\documentclass[aps,prd,12pt,notitlepage,nofootinbib]{revtex4}
    \usepackage{graphicx}
\usepackage{amsmath,amssymb,mathrsfs}
\usepackage{slashed}
\usepackage{epsfig}
\usepackage[font=small,skip=0pt,justification=raggedright]{caption}

\usepackage{ulem}
\usepackage{xcolor}

\newcounter{fig}

\newcommand{\bea}{\begin{eqnarray}}
\newcommand{\eea}{\end{eqnarray}}
\newcommand{\be}{\begin{equation}}
\newcommand{\ee}{\end{equation}}

\newcommand{\re}[1]{(\ref{#1})}

\begin{document}

\title{
Two types of hairy black holes in the\\ (3+1)-dimensional $O(3)$ sigma-model
}
\author{Jutta Kunz}
\email{jutta.kunz@uni-oldenburg.de}
\affiliation{Institute of Physics,
Carl von Ossietzky University Oldenburg,
D-26111 Oldenburg, Germany}

\author{Aliaksei Mikhaliuk}
\email{alekseybsu.mihalyuk@gmail.com}
\affiliation{University of the National Academy of Sciences of Belarus, Minsk 220049, Belarus}

\author{Yakov Shnir}
\email{shnir@theor.jinr.ru}
\affiliation{BLTP, JINR, Dubna 141980, Moscow Region, Russia}

\date{\today}

\begin{abstract}
We construct stationary axially symmetric black holes with scalar hair in the gravitating nonlinear $O(3)$ sigma model with a partly negative symmetry-breaking potential. 
Our results show that it supports two qualitatively different types of asymptotically flat hairy black holes, distinguished by their connection to the corresponding solitonic branches and by the structure of their scalar hair;
the type I black holes are connected to the vacuum branch and type II black holes are connected to regular scalaron-supported spinning solitons. 
The type II black holes have a strongly deformed scalar field structure, with the scalar field taking its vacuum value on the symmetry axis and approaching the anti-vacuum in the near-horizon equatorial region. 
We analyze their global charges, horizon properties, and dependence on the model parameters, comparing the solutions with Kerr black holes. 
For sufficiently weak gravitational coupling, the branch connectivity changes: the fundamental branches of types I and II reconnect to form one continuous family, while their corresponding second branches form another.
We find that, in contrast to the vacuum-connected type I family, the scalaron-supported type II black holes extend to substantially larger horizon radii while preserving a nontrivial scalar configuration outside the horizon.
\end{abstract}

\maketitle

\newpage

%%%%%%%%%%%%%%%%%%%%%%%%%%%%%%%%%%%%%%%%%%%%%%%%%
 \section{Introduction }
  \label{Introduction}

Boson stars provide one of the simplest examples of self-gravitating compact objects supported by scalar fields \cite{Kaup:1968zz,Feinblum:1968nwc,Ruffini:1969qy,Colpi:1986ye,Friedberg:1986tq}.
They arise when a complex scalar field, or equivalently a real multiplet with an internal phase rotation, possesses a harmonic time dependence while the associated stress-energy tensor remains stationary. 
Since the pioneering studies of mini-boson stars and their self-interacting generalizations, these objects have become a standard testing ground for the interplay between gravity, nonlinear field dynamics, and global charges. 
In many models, boson stars exhibit a characteristic branch structure in their mass-frequency diagram, including cusps, spirals, and transitions between stable and unstable sectors (see e.g.~the reviews \cite{Jetzer:1991jr,Lee:1991ax,Schunck:2003kk,Liebling:2012fv,Shnir:2022lba}). 
Rotating boson stars and rotating $Q$-balls add further structure, because the scalar phase winds around the symmetry axis and the solutions carry quantized angular momentum \cite{Kleihaus:2005me}. 
In nonlinear sigma models similar configurations can also occur, although their existence and properties depend sensitively on the target-space constraint and on the scalar self-interaction potential \cite{Verbin:2007fa,Herdeiro:2018djx,Cano:2023bpe,Adam:2025ktm}.

A closely related development is the construction of black holes with synchronized scalar hair \cite{Herdeiro:2014goa,Herdeiro:2015gia,Herdeiro:2015waa}.
In these solutions the scalar field is not static, but its harmonic dependence is compatible with a stationary and axisymmetric geometry because the scalar frequency is locked to the angular velocity of the horizon. 
The corresponding synchronization condition ensures the absence of scalar flux through the event horizon and allows a nontrivial scalar condensate to remain in equilibrium outside the black hole. 
This mechanism underlies the existence of rotating hairy black holes connected to boson-star families and provides a way to evade the assumptions of standard no-hair theorems \cite{Herdeiro:2015waa}. 
It has been realized in several models, including black holes with synchronized hair in the four-dimensional $O(3)$ sigma model and in self-interacting scalar theories \cite{Herdeiro:2018djx,Kunz:2019sgn}. 
Such configurations serve as an important bridge between regular solitonic compact objects and black holes endowed with matter hair.

Scalar theories with self-interaction potentials that are not everywhere positive provide another mechanism for supporting nontrivial self-gravitating configurations. 
Such potentials may relax some of the usual assumptions entering no-hair  arguments and can support solitonic and black-hole solutions with properties that differ substantially from those of models with positive-definite matter potentials \cite{Nucamendi:1995ex,Bechmann:1995sa,Kleihaus:2013tba,Dzhunushaliev:2008bq,Chew:2023olq}.

The nonlinear $O(3)$ sigma model with a partly negative symmetry-breaking potential provides a particularly interesting realization of this mechanism.
The scalar triplet is constrained to the unit sphere, while the potential contains a negative higher-order contribution in addition to the usual positive pion-mass term. 
This interaction supports non-topological scalaron solutions in flat space \cite{Ferreira:2025xey,Mikhaliuk:2026tdt}. 
When gravity is included, a corresponding scalaron-supported branch of spherically symmetric boson stars arises \cite{Kunz:2025tsy}. 
Moreover, since the potential relaxes assumptions entering no-hair theorems for self-gravitating nonlinear sigma models \cite{Heusler:1992ss}, it is interesting to investigate whether the scalaron-supported sector persists for stationary spinning configurations and in the presence of an event horizon.

In the present work we first construct regular axially symmetric spinning solitonic solutions of this model. 
We find two qualitatively different regular families with azimuthal winding $n=1$. 
The type I branch is connected to vacuum fluctuations in the limit $\omega\to\mu$, whereas the type II branch belongs to the scalaron-supported sector associated with the partly negative potential.
Both families satisfy the regularity condition that the scalar field takes its vacuum value on the symmetry axis, but they possess distinctly different scalar field structure.

We then construct the corresponding stationary axially symmetric black holes with scalar hair, where the harmonic dependence of the scalar triplet is synchronized with the angular velocity of the event horizon. 
For small horizon radius the two types of black holes can be connected to the corresponding regular spinning solitons. 
The type I black holes constitute the vacuum-connected family, whereas the type II black holes form the scalaron-supported family. 
A characteristic feature of the latter is a strongly deformed scalar distribution: the field takes its vacuum value on the symmetry axis while it may approach the anti-vacuum in the near-horizon equatorial region. 
Furthermore, the type II family extends to substantially larger horizon radii than the type I family.
Analogously to the globally regular solutions, we find that decreasing the gravitational coupling leads to a bifurcation and subsequent reconnection of the black-hole branches. 
The fundamental branches of types I and II form one continuous family, while their corresponding second branches reconnect into a separate family that emerges from Kerr and may approach Kerr-like global properties without losing its nontrivial scalar hair.
We analyze the domains of existence and branch structure of both families as functions of the gravitational coupling, the potential parameter, the horizon radius, and the scalar frequency, together with their global charges and horizon properties, and compare the solutions with the corresponding Kerr black holes.

%%%%%%%%%%%%%%%%%%%%%%%%%%%%%%%%%%%%%%%%%%%%%%%%%%

\section{The Model}
%%%%%%%%%%%%%%%%%%%%%%%%%%%%%%%%%%
\subsection{The action and field equations}

We consider the nonlinear $O(3)$ sigma model minimally coupled to Einstein's gravity in an asymptotically flat (3+1)-dimensional space-time. 
The action of the model is
\be
S=\int d^4 x \sqrt{ -g }
\left(\frac{R}{16\pi G} - L_{m}\right)\, .
\label{lag}
\ee
Here the gravity part is the usual Einstein-Hilbert action, $R$ is the Ricci scalar associated with the spacetime metric $g_{\mu\nu}$ with the determinant ${g}$, and $G$ is Newton's constant.
The Lagrangian of the matter fields $L_m$ is given by the nonlinear $O(3)$ sigma-model
\be
L_{m} = \frac12 (\partial_\mu\phi^a)^2 + U(\phi) \, .
\label{lag_O3}
\ee
Here the real triplet of the scalar fields $\phi^a$, $a=1,2,3$, is restricted to the surface of the unit sphere, $(\phi^a)^2=1$, and $U(\phi)$ is a symmetry breaking potential \cite{Leese:1989gi,Salmi:2014hsa,Gillard:2015eia,Samoilenka:2015bsf,Ferreira:2025xey,Mikhaliuk:2026tdt},
\be
U(\phi)=\mu^2 (1 -\phi^3) - \beta (1-\phi^3)^4 \,  .
\label{pot-mod}
\ee
Here $\mu>0$ sets the mass scale and $\beta$ is a real parameter. 
Importantly, the potential \re{pot-mod}, in addition to the usual weakly attractive term, also contains a higher-order term which, with the opposite sign, can give rise to short-range repulsion between solitons in related models \cite{Leese:1989gi,Salmi:2014hsa,Gillard:2015eia,Samoilenka:2015bsf}.

For sufficiently large positive $\beta$, the higher-order contribution allows the potential to become negative in part of the target space; below we focus mainly on this regime. 
The potential explicitly breaks the $O(3)$ symmetry while preserving rotations in the $(\phi^1,\phi^2)$ plane. 
Thus, the Lagrangian (\ref{lag_O3}) remains invariant under the global $U(1)$ transformations
\begin{equation}
 (\phi^1+i\phi^2)\rightarrow e^{i\gamma}(\phi^1+i\phi^2),
\end{equation}
where $\gamma$ is a constant phase.
Hence,
\be
j^\mu=\phi^1 \partial^\mu\phi^2 - \phi^2 \partial^\mu\phi^1
\label{ncurrent}
\ee
is the conserved Noether current.
The corresponding Noether charge is
\be
Q=-\int_{\Sigma} j^\mu n_\mu\,d\Sigma ,
\ee
where $n^\mu$ is the future-directed unit normal to the spacelike hypersurface $\Sigma$.

Variation of the action \re{lag} with respect to the metric leads to the Einstein equations
\be
R_{\mu\nu}-\frac{1}{2}Rg_{\mu\nu}
=\frac{\alpha^2}{2}T_{\mu\nu}
%R_{\mu\nu} -\frac12 R g_{\mu\nu} = 2\alpha^2 T_{\mu\nu} \, ,
\label{einsteq}
\ee
where the  scalar energy-momentum tensor is
\be
T_{\mu\nu} = \partial_\mu\phi^a \partial_\nu\phi^a
 - g_{\mu\nu}\left[\frac{g^{\rho\sigma}}{2}  \partial_\rho\phi^a
\partial_\sigma\phi^a +  U(\phi)\right] \, .
\label{Teng}
\ee
The corresponding scalar field equations are
\begin{equation}
\nabla_\mu\nabla^\mu\phi^a
+\phi^a\left(\nabla_\mu\phi^b\nabla^\mu\phi^b\right)
-\left(\delta^{ab}-\phi^a\phi^b\right)
\frac{\partial U}{\partial\phi^b}=0 .
\label{eqfield}
\end{equation}
Here the projection onto the tangent space of the target sphere
implements the constraint $\phi^a\phi^a=1$.

Rescaling of the coordinates $x_\nu\to \mu x_\nu$ and the parameter $\beta \to \beta/\mu^2$ allows us to set $\mu=1$, the effective gravitational coupling becomes $\alpha^2=16\pi G\mu^2$.

\subsection{Stationary axially symmetric ansatz and boundary conditions}

We are interested in stationary spinning asymptotically flat axially symmetric solutions of the system \re{einsteq}, \re{eqfield} with and without an event horizon.
To construct such solutions numerically we employ the line element in Lewis-Papapetrou form
\cite{Kleihaus:2000kg,Herdeiro:2018djx,Herdeiro:2015gia}
\be
ds^2=g_{\mu\nu}dx^\mu dx^\nu= -F_0 dt^2 +F_1(dr^2 + r^2 d\theta^2 ) + r^2 \sin^2 \theta \,
F_2 \left(d\varphi - W dt \right)^2 \, ,
\label{metric}
\ee
where the four metric functions $F_0,\, F_1,\, F_2$ and $W$ depend on the isotropic radial coordinate
$r$ and polar angle $\theta$ only.
For the scalar $O(3)$ field, we adopt the ``trigonometric" stationary ansatz \cite{Ferreira:2025xey}
\be
    \phi^a=[\sin{f(r,\theta)}\cos{(n\varphi-\omega t)},\text{ }
    \sin{f(r,\theta)}\sin{(n\varphi-\omega t)},\text{ }\cos{f(r,\theta)}] \, ,
    \label{anssfield}
\ee
where $n\in \mathbb{Z}$ is the azimuthal harmonic index.

We are looking for asymptotically flat stationary solutions of two kinds: globally regular spinning solitons and black holes with scalar hair. 
The two classes share the same boundary conditions at spatial infinity, on the symmetry axis, and on the equatorial plane, but differ at the inner radial boundary. 
For the black-hole solutions, we assume the existence of a rotating event horizon located at a constant value of the radial coordinate $r=r_h>0$. 
The Killing vector generating the horizon is
\be
\chi = \xi+ \Omega_h \eta\, ,
\label{Killing}
\ee
where $\xi,~\eta$ are two commuting Killing vector fields,
$\xi=\partial_t$ and $\eta=\partial_\varphi$.

The corresponding boundary conditions are found by considering the asymptotic expansions of the equations on the boundary of the domain of integration together with the assumption of regularity and asymptotic flatness. 
In particular, a power series expansion near the horizon implies that
$W\left.\right|_{r=r_h}=\Omega_h = \omega/n$ where the quantity
$$
\Omega_h = -\frac{g_{\varphi t}}{g_{\varphi\varphi}}\left.\right|_{r=r_h}
$$
is the horizon angular velocity. 
The corresponding synchronisation condition between the field frequency and the event horizon angular velocity $\omega=n \Omega_h$  implies that there is no net flux of the scalar field across the horizon and also that the components $T_{tt}$ and $T_{t\varphi}$ of the energy-momentum tensor \re{Teng} are finite at the horizon.

It is convenient to employ the following exponential parametrization of the metric functions for both classes of solutions,
\be
F_0=\frac{\left(1-\frac{r_h}{r}\right)^2}{\left(1+\frac{r_h}{r}\right)^2}\, e^{2f_0},\quad
F_1=\left(1+\frac{r_h}{r}\right)^4\, e^{2f_1}, \quad
F_2=\left(1+\frac{r_h}{r}\right)^4\, e^{2f_2}, \,
\label{F012}
\ee
For the globally regular solutions we set $r_h=0$, so that the radial domain is $r\in[0,\infty)$. 
Regularity at the origin requires the metric functions and $W$ to have vanishing radial derivatives there. 
For the regular $n=1$ solutions considered in this work, the scalar field takes its vacuum value at the origin. 
Hence, the boundary conditions at $r=0$ are
\begin{equation}
\left.f\right|_{r=0}=0,\qquad
\left.\partial_r f_0\right|_{r=0}
=\left.\partial_r f_1\right|_{r=0}
=\left.\partial_r f_2\right|_{r=0}
=\left.\partial_r W\right|_{r=0}=0 .
\end{equation}
For the black-hole solutions, $r_h>0$ and the inner radial boundary is
the event horizon.
For the $n=1$ solutions considered below, the boundary conditions at the
horizon become
\begin{itemize}
    \item at $r=r_h:\qquad \partial_r f = \partial_r f_0=\partial_r f_1 =
    \partial_r f_2 = 0 \ ,
    W=\omega$ \, .
\end{itemize}
Asymptotic flatness requires that, as $r\to \infty$, the metric approaches the flat space limit, and the $O(3)$ scalar field takes its vacuum value,
\begin{itemize}
\item at $r\to \infty:\qquad  f = 0,\quad  f_0= f_1 =
    f_2 = W= 0$ \, .
\end{itemize}
Axial symmetry and regularity of the $n=1$ ansatz then impose the following boundary conditions on the symmetry axis for both the globally regular and
black-hole solutions.
\begin{itemize}
\item at $\theta\to 0,\pi: \qquad  f = 0,\quad  \partial_\theta
f_0= \partial_\theta f_1 =
   \partial_\theta f_2 =\partial_\theta W= 0$ \, .
\end{itemize}

We also require the solutions to be symmetric with respect to reflection $\theta \to \pi -\theta$ in the equatorial plane $\theta=\pi/2$. 
It allows us to restrict the range of values of the angular variable as $\theta \in [0,\pi/2]$.
The corresponding boundary conditions on the equatorial plane are
\begin{itemize}
\item at $\theta\to \pi/2: \qquad  \partial_\theta f = 0,\quad  \partial_\theta
f_0= \partial_\theta f_1 =
   \partial_\theta f_2 =\partial_\theta W= 0$ \, .
\end{itemize}
Further, the absence of a conical singularity at the symmetry axis requires that the deficit angle should vanish, $\delta =
2\pi\left(1-\lim_{\theta\to0}
\sqrt{\frac{F_2}{F_1}}\right)=0$.   
Hence, the solutions should satisfy the constraint $F_1\left.\right|_\theta=0 = F_2\left.\right|_\theta=0$. 
In our numerical scheme we explicitly imposed this restriction.

\subsection{Quantities of interest}

For the black-hole solutions, the horizon quantities are determined by
the metric functions at $r=r_h$.
We first consider the Hawking temperature $T_h$, which is determined by the surface gravity $\kappa^2=-\frac12
\nabla_\mu \chi_\nu \nabla^\mu \chi^\nu$,
\be
T_h=\frac{\kappa}{2\pi}=\frac{1}{16\pi r_h}\exp\left[(f_0-f_1)\left.\right|_{r=r_h} \right] .
\ee
The horizon area is
\be
A_h=32 \pi r_h^2 \int\limits_0^\pi d\theta \sin \theta \exp\left[(f_1+f_2)\left.\right|_{r=r_h} \right] \, .
\ee
The black hole entropy is $S=A_h/(4G)$.

The measure of the horizon deformation is the ratio of polar to equatorial circumferences $\epsilon=L_p/L_e$, where, with the metric \re{metric},
\be
L_p=2 r_h\int\limits_0^\pi d \theta \sqrt{F_1(r_h,\theta)} , \qquad
L_e=2\pi r_h \sqrt{F_2(r_h, \pi/2)} \, .
\ee

For both the globally regular spinning solitons and the hairy black holes, the asymptotic expansion of the metric functions at spatial infinity yields the total ADM mass $M$ and angular momentum $J$,
\be
g_{tt}=-1+\frac{2GM}{r}+ {\cal O}(r^{-2}), \qquad g_{\varphi t}= - \frac{2 G J}{r}\sin^2 \theta + {\cal O}(r^{-2})
\label{expMJ}
\ee

For the black-hole solutions, the ADM charges can be decomposed into a horizon contribution and a bulk scalar-field contribution, $M=M_h+M_\phi$ and $J=J_h+J_\phi$,
\be
\begin{split}
M_h&=-\frac{1}{8\pi G} \oint_S dS_{\mu\nu}\nabla^\mu \xi^\nu\, , \qquad
J_h= \frac{1}{16 \pi G}\oint_S dS_{\mu\nu} \nabla^\mu \eta^\nu\, ;\\
M_\phi&=- \int_V dS_\mu (2 T^\mu_\nu \xi^\nu - T \xi^\mu)\, , \qquad
J_\phi=\int_V dS_\mu \left(T_\nu^\mu \eta^\nu - \frac12 T \eta^\mu\right)
\label{MJ}
\end{split}
\ee
where $S$ is the horizon 2-sphere and $V$ is an asymptotically flat spacelike hypersurface bounded by the horizon.
It's useful to introduce a ``hairiness'' $h$, which allows one to trace evolution of solutions from a boson star $(h=1)$ to a Kerr black hole $(h=0)$: $h=M_\phi/M$.

The physical quantities of the hairy black hole are related via the first law of thermodynamics of black holes,
\be
dM = T_h dS +\Omega_h dJ
\ee
and the corresponding Smarr relation
\be
M=2T_h S + 2\Omega_h J_h + M_\phi
\label{Smarr}
\ee

To facilitate the numerical treatment of the semi-infinite radial domain, we introduce the compactified coordinate $x = \frac{r-r_h}{r+C}$, which maps the semi-infinite region $[r_h;\infty)$ onto the unit interval $[0; 1]$. 
For black holes, $x=0$ corresponds to the event horizon $r=r_h$.
For globally regular solutions we set $r_h=0$, so that $x=r/(r+C)$ and $x=0$ corresponds to the regular origin. 
In both cases spatial infinity is mapped to $x=1$.
$C$ is an arbitrary constant used to adjust the contraction of the grid.
With this compactification, the asymptotic expansion (\ref{expMJ}) gives $GM=2r_h+(C+r_h) \left.\frac{\partial f_0}{\partial x}\right|_{x=1}$, where the first term originates from the explicit Schwarzschild-like factor in $F_0$ in Eq.~(\ref{F012}).

The black-hole solutions are characterized by four continuous parameters $(\alpha,\beta,r_h,\omega)$ and the discrete winding number $n$. 
For the globally regular solutions one sets $r_h=0$, leaving $(\alpha,\beta,\omega)$ and $n$ as the relevant parameters.
Within this parameter space we find two distinct regular families of spinning solitons and their corresponding hairy black-hole solutions.
Notably, we do not find spherically symmetric ($n=0$) hairy black holes of either type.
To simplify our analysis, we restrict our consideration to the spinning parity-even solutions with winding number $n=1$ and we will mainly consider solutions with some particular values of the parameter $\beta$, typically we take $\beta=0.5$.
In our analysis, we restrict ourselves to a few particular values of the effective gravitational coupling $\alpha = 0.6$, $0.5$, $0.4$, and $\alpha = 0.3$, in order to illustrate the change of the branch structure at weaker gravitational coupling.

\section{Spinning $O(3)$ boson stars}

First, we consider regular axially symmetric spinning solitonic solutions of the model \re{lag} with the symmetry-breaking potential \re{pot-mod}. 
The structure of these configurations reflects the interplay between gravitational attraction, rotation, and the short-range scalar
self-interaction.
We obtain two regular classes of $n=1$ solutions, analogous to the case of the $n=0$ solutions.
The type I boson stars are connected to vacuum fluctuations in the Newtonian limit, whereas the type II boson stars are associated with static nontopological self-gravitating solitons supported by the partly negative potential. 
For both regular $n=1$ branches the scalar field takes the vacuum value on the symmetry axis and at spatial infinity.

\begin{figure}[t!]
\begin{center}
\includegraphics[height=.34\textheight,  angle =-90]{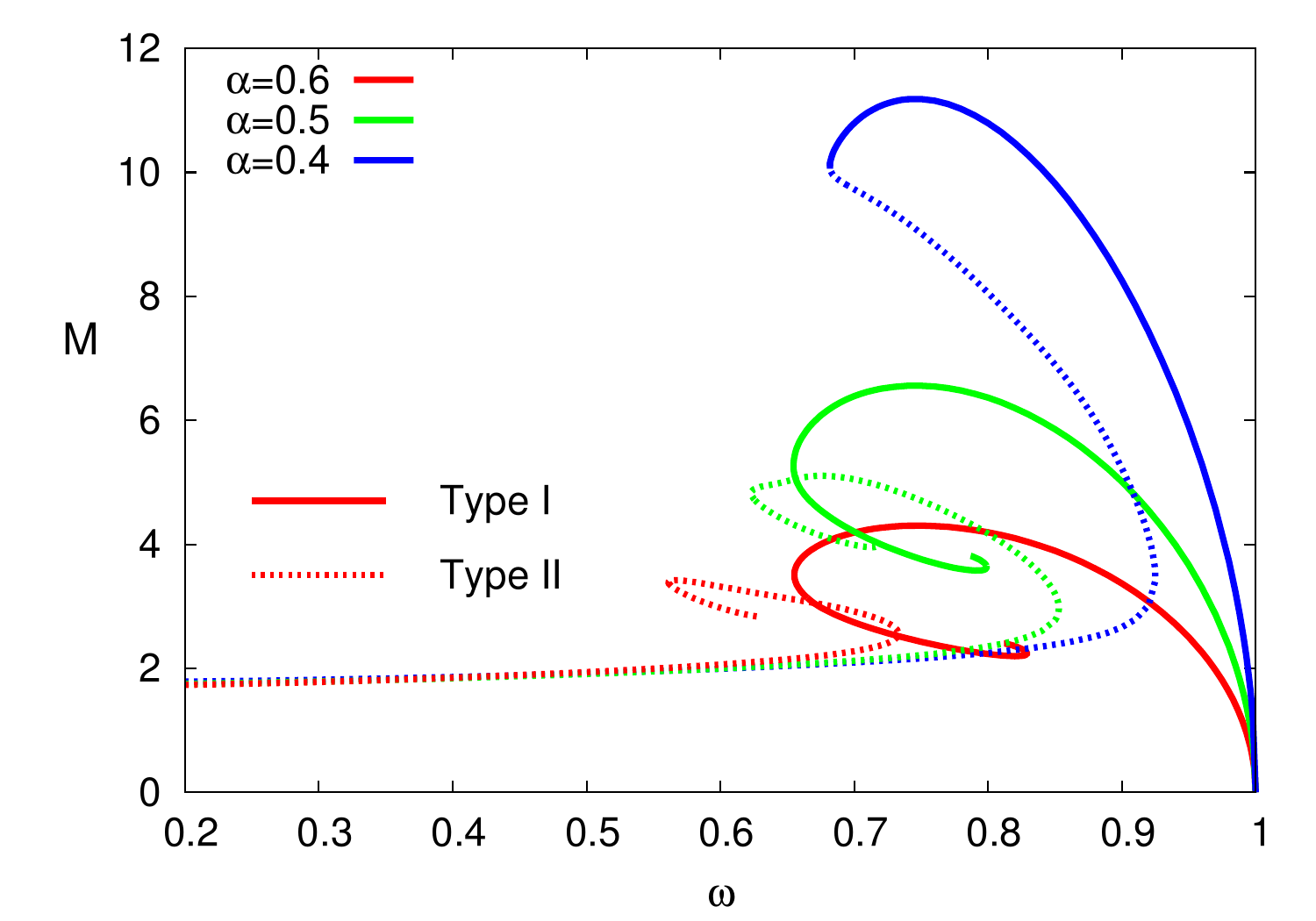}
\includegraphics[height=.34\textheight,  angle =-90]{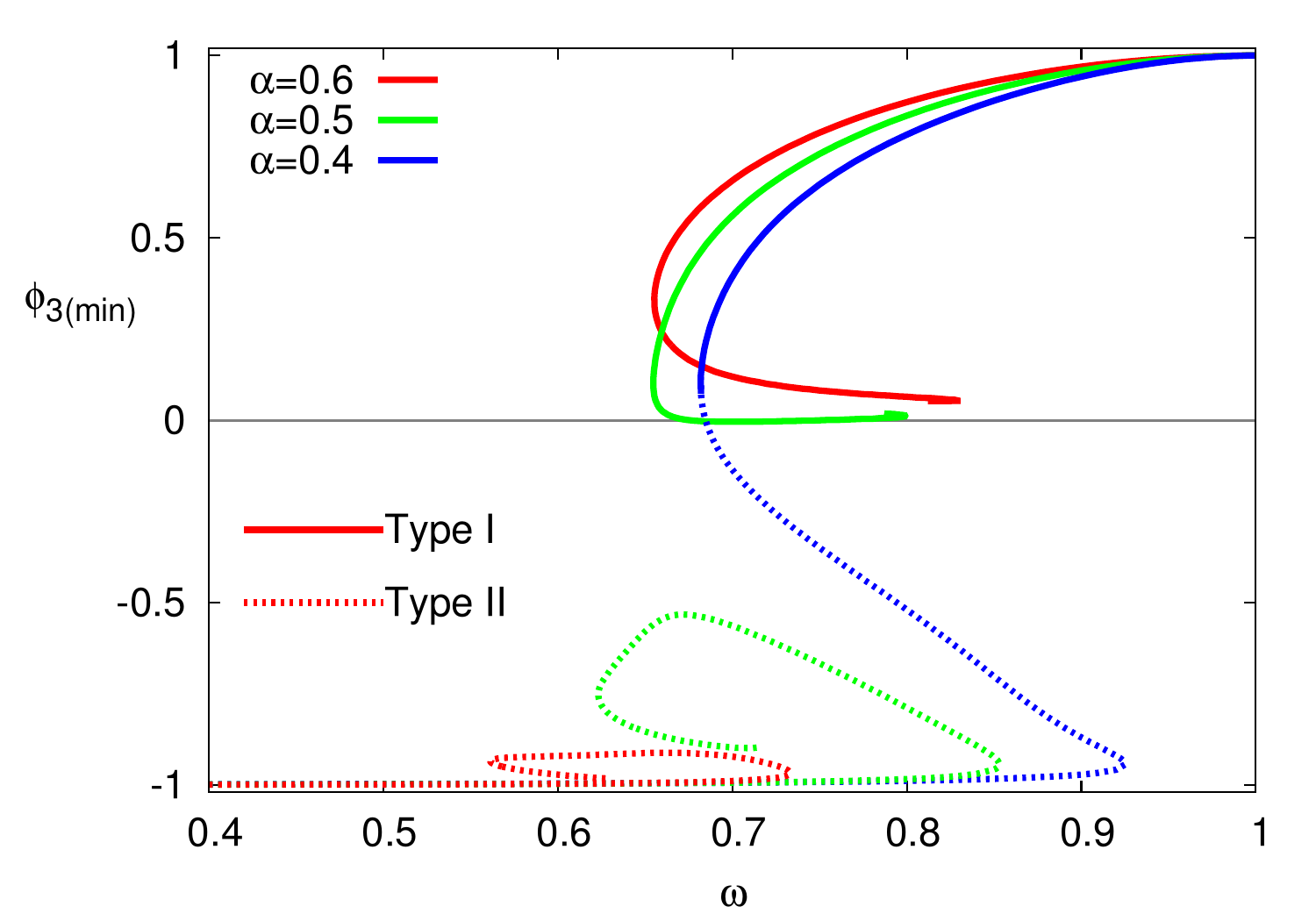}
\includegraphics[height=.34\textheight,  angle =-90]{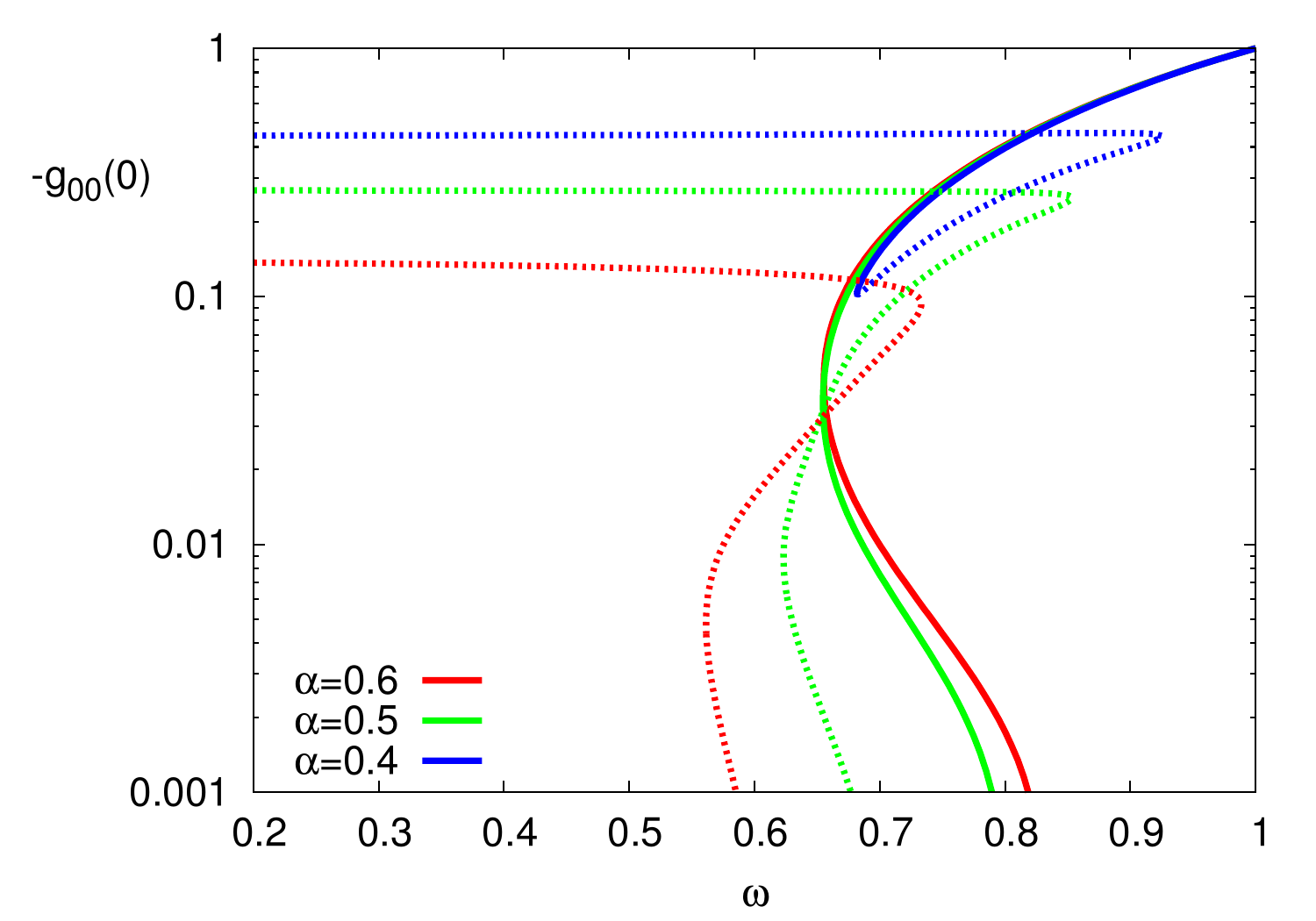}
\end{center}
\caption{\small
The $O(3)$ spinning boson stars of the types I and II:
the ADM mass $M$ in units of $16\pi$ (upper left plot) and the minimum value of the scalar field component $\phi_3$ (upper right plot) and the central value of the metric component $-g_{00}$ (bottom plot) are displayed vs the angular frequency $\omega$ for $\beta=0.5$ and a set of values of the gravitational coupling $\alpha$. 
The solid and dashed lines correspond to solutions of types I and II, respectively.}
    \label{figBS1}
\end{figure}
%%%%%%%%%%%%%%%%%%%%%%%%%%%%%%%%%%%%%%%%%%%%%%%%%%%%%%%%%%%%%%%%%%%%%

Spinning type I boson stars arise smoothly as linearized perturbations of the vacuum $\phi_0=(0,0,1)$ as the angular frequency decreases below the mass threshold $\omega_{max}=\mu =1$ in rescaled units. 
If the gravitational coupling is relatively large, $\alpha \ge 0.46$, a characteristic spiral-like dependence of the configuration's charge and mass on the frequency is observed (see Fig.~\ref{figBS1}, left plot). 

For the type I solutions, the presence of a negative-energy contribution from the symmetry breaking potential \re{pot-mod} has practically no effect on the well-known scenario of spiral evolution of boson stars. 
However, there is a significant difference between such solutions and ordinary boson stars. 
For the $O(3)$ configurations, the profile function is bounded, $0\leq f\leq\pi$. 
During the evolution along the spiral, its maximal value $f_{\rm max}$ increases from the vacuum value $f=0$ towards some limiting value. 
Since $\phi_3=\cos f$, this corresponds to the minimum value $\phi_{3 (\text{min})}=\cos f_{\rm max}$ decreasing from the vacuum $\phi_3=1$ towards a limiting value, as seen in Fig.~\ref{figBS1}, right plot.
For the type I boson stars this limiting value remains positive.

Type II boson stars are associated with a static self-gravitating soliton; they exist due to the presence of the partially negative self-interaction potential \re{pot-mod}.
These solutions reside in the frequency range $\omega \in [0,\omega_{II}^{max} < 1]$.
For the  type II boson stars the minimum value $\phi_{3 (\text{min})}=\cos f_{\rm max}$ of the scalar component $\phi_3$ deviates from \textit{anti-vacuum} $\phi_3=-1$ as the angular frequency increases from zero. 
The magnitude of the deviation increases as the gravitational coupling decreases. 
Similar to the case of the usual boson stars, the metric function $g_{00}$ at the center of the configurations of the types I and II approaches zero, as seen in Fig.~\ref{figBS1}, bottom plot.
Both regular boson-star branches exhibit the typical spiraling and oscillating pattern with successive backbendings, while the mass and the charge tend to finite limiting values at the centers of the corresponding spirals. 

The situation changes dramatically as the gravitational coupling approaches a critical value, $\alpha < 0.46$ for $\beta=0.5$.
At this value the two regular branches bifurcate and reconnect, as seen in the mass-frequency diagram in Fig.~\ref{figBS1}. 
Thus, there results a connected set of regular self-gravitating solitonic solutions of the $O(3)$ sigma model, which exists for the whole range of values of the frequency $\omega \in [0,1]$. 
In addition, we expect a set of spirals that are detached from the main branches, analogous to the case of the non-rotating solutions \cite{Kunz:2025tsy}.

%%%%%%%%%%%%%%%%%%%%%%%%%%%%%%%%%%%%%%%%%%%%%%%%%%%%%%%%%%%%%%%%%%%%%

\section{Hairy $O(3)$ black holes}

%%%%%%%%%%%%%%%%%%%%%%%%%%%%%%%%%%%%%%%%%%%%%%%%%%%%%%%%%%%%%%%%%%%%%%
\subsection{Black holes with type I hair}

\begin{figure}[t!]
\begin{center}
\includegraphics[height=.24\textheight,  angle =0]{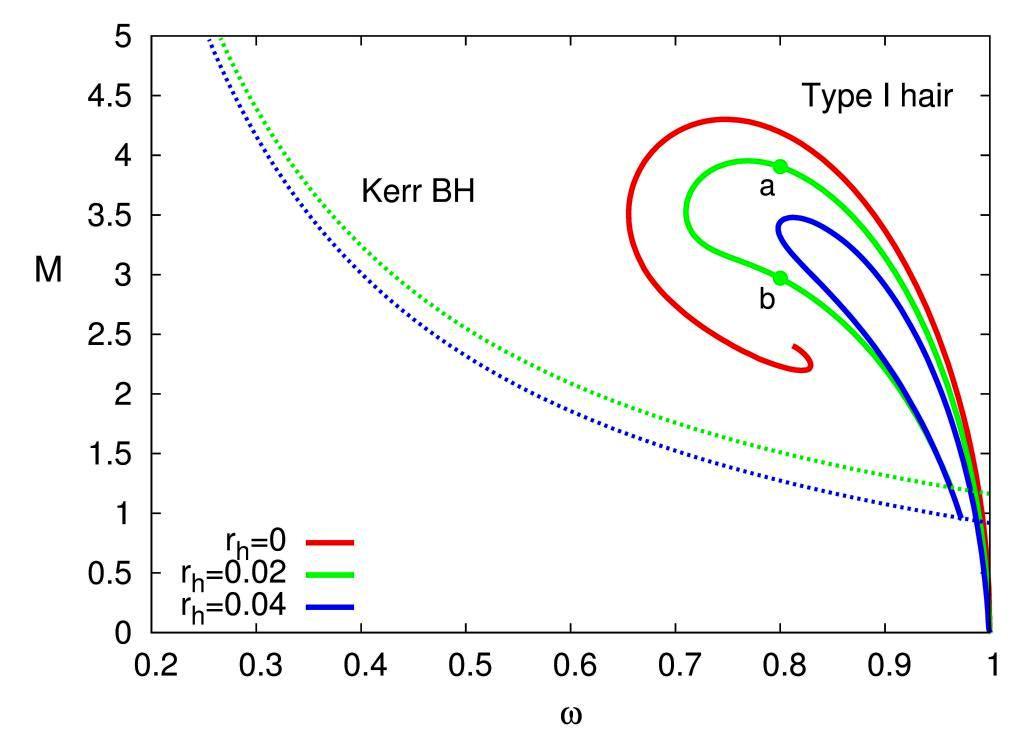}
\includegraphics[height=.24\textheight,  angle =0]{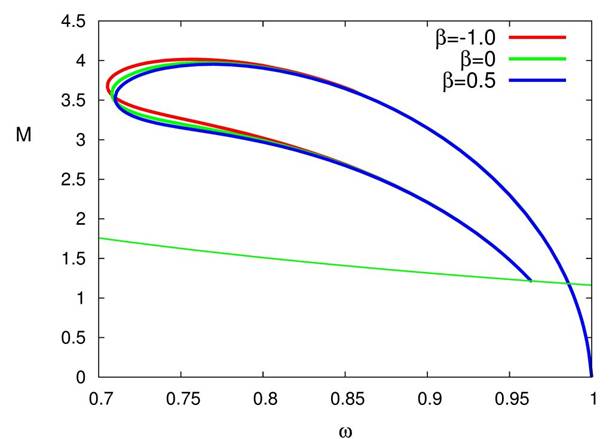}
\end{center}
\caption{\small
The $O(3)$ black holes with type I hair:
ADM mass $M$ in units of $16\pi$ is displayed vs the angular frequency $\omega$ for $\beta=0.5$ and a set of values of the horizon radius $r_h$ (left plot), and for a set of values of the parameter $\beta$ for $r_h=0.02$ (right plot), both for $\alpha=0.6$.
For comparison the corresponding vacuum Kerr solutions are also shown (dotted lines).}
    \label{figBH1}
\end{figure}

The $O(3)$ black holes with type I hair are qualitatively similar to the perturbative $O(3)$ hairy black holes discussed in \cite{Herdeiro:2018djx}. 
Their properties weakly depend on the value of the potential parameter $\beta$, which can also be negative, see Fig.~\ref{figBH1}, right plot. 
The left plot of Fig.~\ref{figBH1} exhibits the variation of the ADM mass of the spinning hairy BHs, at a fixed value of the gravitational coupling $\alpha=0.6$ and the parameter $\beta=0.5$, versus the angular frequency $\omega$, for a set of values of the horizon radius $r_h$. 
The solutions are smoothly connected to the spinning boson stars in the limit $r_h\to 0$. 
Both the $O(3)$ boson stars and type I hairy black holes arise as the angular frequency decreases from the mass threshold.
The ADM mass of all solutions increases and approaches a maximum on the first (fundamental) branch at a critical value of the frequency $\omega_{cr}^{I}$. 
The maximum mass decreases as the horizon radius increases.     

%%%%%%%%%%%%%%%%%%%%%%%%%%%%%%%%%%%%%%%%%%%%%%%%%%%%%%%%%%%%%%%%%%%%%

%%%%%%%%%%%%%%%%%%%%%%%%%%%%%%%%%%%%%%%
\begin{figure}[t!]
\begin{center}
\includegraphics[height=.24\textheight,  angle =-0]{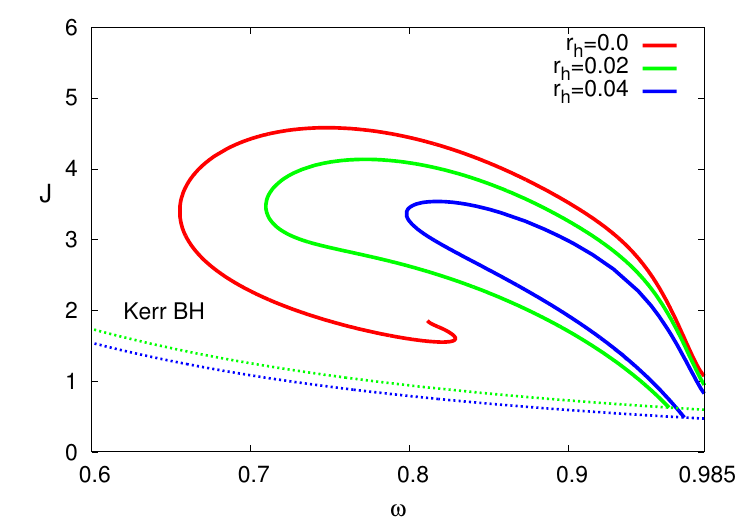}
\includegraphics[height=.24\textheight,  angle =-0]{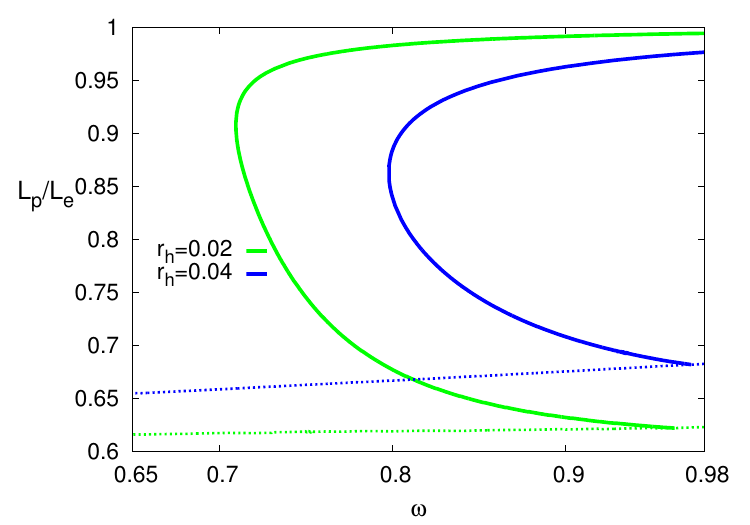}
\includegraphics[height=.24\textheight,  angle =-0]{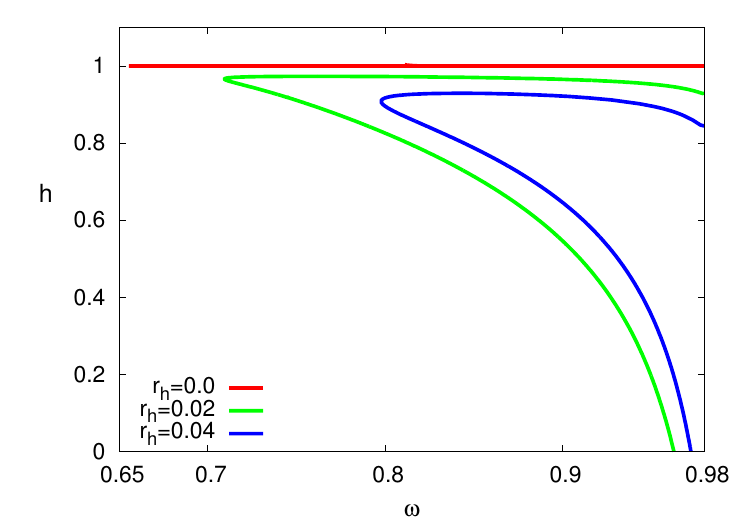}
\end{center}
\caption{\small
The $O(3)$ black holes with type I hair:
Angular momentum $J$ in units of $16\pi$ (left plot), ratio of polar to equatorial circumferences (right plot) and ``hairiness'' $h$ (bottom plot) are displayed vs the angular frequency $\omega$ for a set of values of the horizon radius $r_h$ at $\beta=0.5$, $\alpha=0.6$. 
}
    \label{fig_JvsW_typeI}
\end{figure}
%%%%%%%%%%%%%%%%%%%%%%%%%%%%%%%%%%%%%%%%%%%%%%%%%%%%%%%%%%%%%%%%%%%%%

%%%%%%%%%%%%%%%%%%%%%%%%%%%%%%%%%%%%%%%
\begin{figure}[t!]
\begin{center}
\includegraphics[height=.20\textheight,  angle =-0]{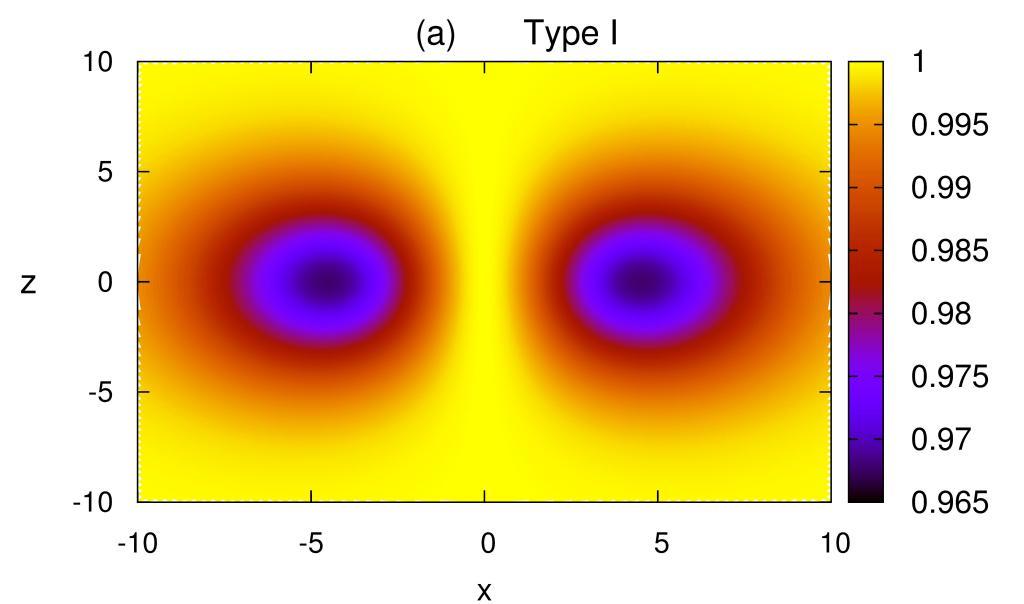}
\includegraphics[height=.20\textheight,  angle =-0]{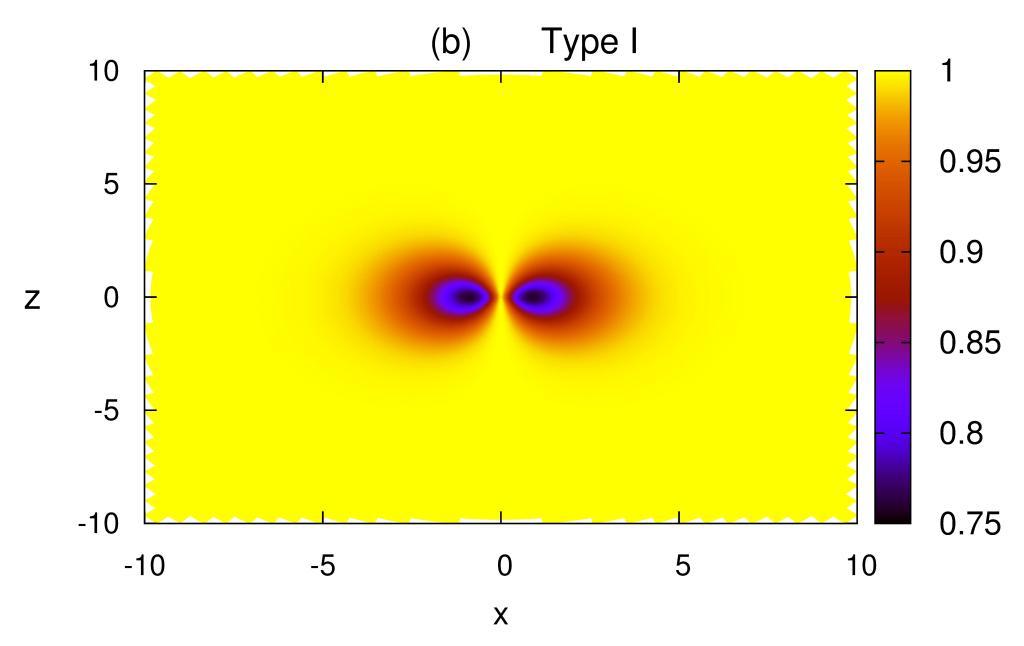}
\end{center}
\caption{\small
The type I $O(3)$ hairy black holes with small event horizon:
Contour plots of the scalar field component $\phi_3$ in the $x z$ plane are displayed for the configurations, indicated by the corresponding dots in the left panel of Fig.~\ref{figBH1}, for $\beta=0.5$ and the horizon radius $r_h=0.02$. 
}
    \label{fig_contour}
\end{figure}
%%%%%%%%%%%%%%%%%%%%%%%%%%%%%%%%%%%%%%%%%%%%%%%%%%%%%%%%%%%%%%%%%%%%%

For most non-zero values of the horizon radius a two-branch scenario is observed\footnote{The multi-branch structure appears to exist \cite{Herdeiro:2018djx} at very small values of $r_h$  for all values of $\beta$.}. 
A turning point arises at a minimum value of the frequency $\omega_{min}^{I}$ as the fundamental branch continues as the second (lower, forward) branch, ending on the Kerr solution as $\omega \to \omega_{end}^I$. 
The minimal frequency $\omega_{min}^{I} $ increases slightly with decreasing $\beta$, as seen in Fig.~\ref{figBH1}, right plot. 
Also, the maximum value of the frequency increases as $r_h$ increases.
It approaches $\omega_{end}^I$ at a maximal value of the horizon radius. 
The corresponding angular momentum $J$, horizon deformation $L_p/L_e$ and hairiness $h$ are shown in Fig.~\ref{fig_JvsW_typeI}. 
The angular momentum follows the same two-branch pattern as the ADM mass. 
Along the second branch the scalar-field contribution decreases and the solutions approach the Kerr limit, as reflected by the decrease of the hairiness parameter $h$.

For all solutions of the first type, the component of the scalar field $\phi_3$ differs relatively little from the vacuum value $\phi_3=1$, as seen in Fig.~\ref{fig_contour}. 
Figure~\ref{fig_contour} illustrates the scalar-field distribution for two representative type I configurations on the two branches.
The scalar field distribution is a typical toroidal pattern for rotating configurations, with a clear maximum in the equatorial plane and a suppression of the profile function at high latitudes. 
The profile function vanishes along the symmetry axis.
On the second (forward) branch the solutions become more compact.

\begin{figure}[t!]
    \centering
    \includegraphics[width=0.33\linewidth, angle = -90]{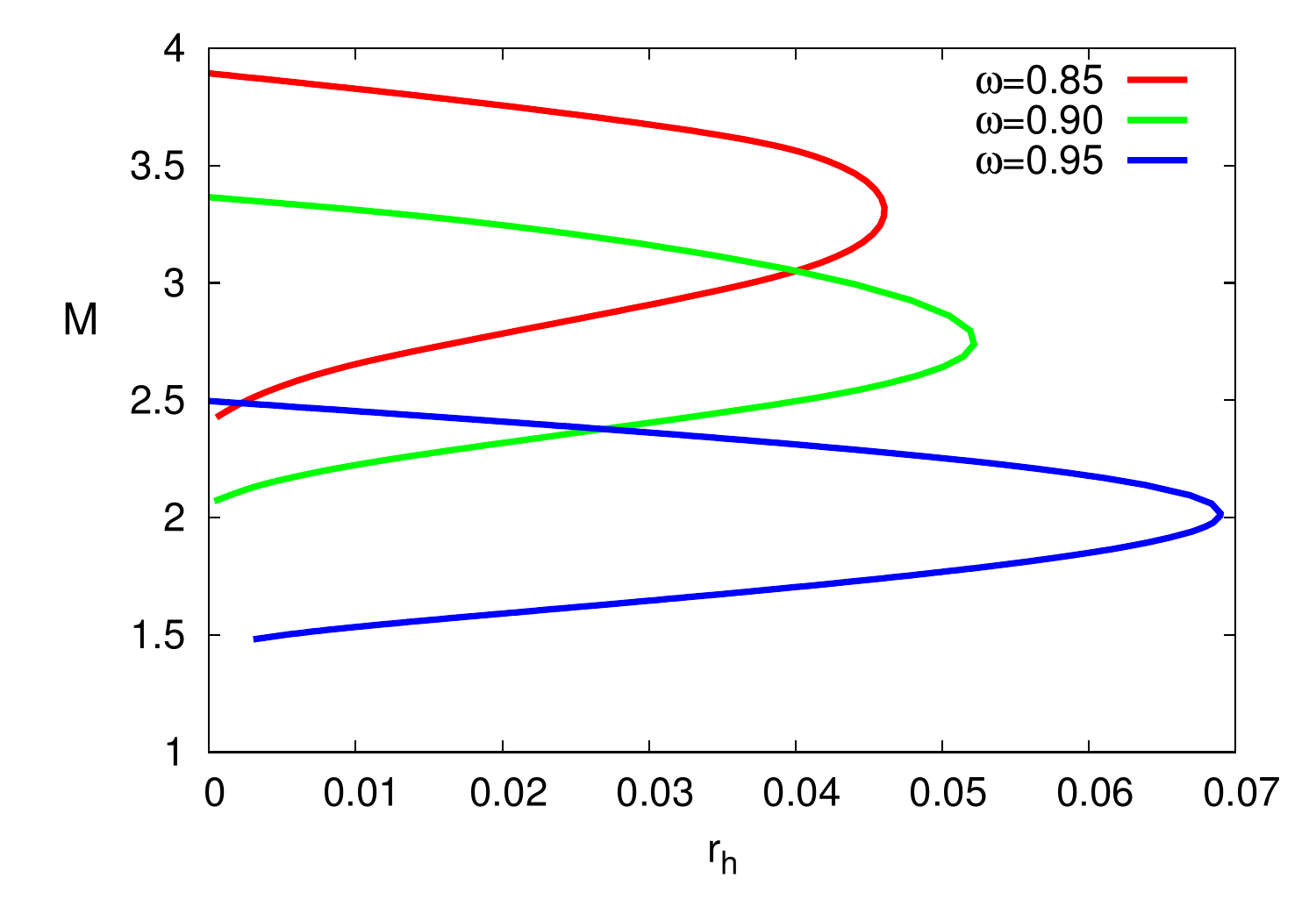}
    \includegraphics[width=0.33\linewidth, angle = -90]{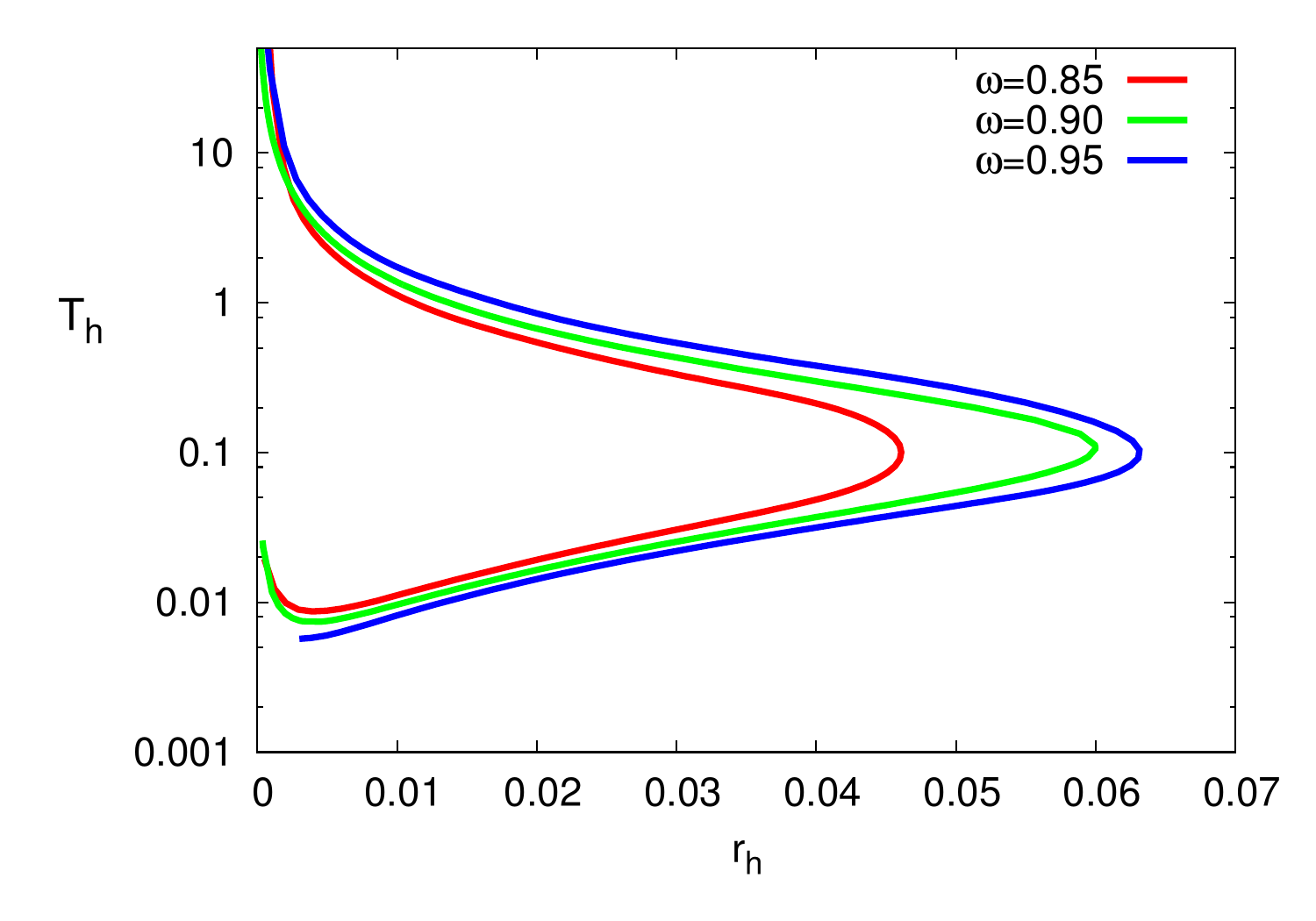}
    \caption{The type I $O(3)$ black hole with small event horizon: 
    ADM mass $M$ in units of $16\pi$ (left) and the Hawking temperature $T_h$ (right) vs the horizon radius $r_h$ for different values of $\omega$, $\beta=0.35$, $\alpha=0.6$.}
    \label{fig_type1_MTvsRh}
\end{figure}

The dependence of the ADM mass and the Hawking temperature on the horizon radius is shown in Fig.~\ref{fig_type1_MTvsRh} for three values of $\omega$.
The ADM mass is larger for smaller values of $\omega$.
Along the second branch, the mass decreases, whereas the Hawking temperature increases.

%%%%%%%%%%%%%%%%%%%%%%%%%%%%%%%%%%%%%%%%%%%%%%%%%%%%%%%%%%%%%%%%%%%%%
%%%%%%%%%%%%%%%%%%%%%%%%%%%%%%%%%%%%%%%%%%%%%%%%%%%%%%%%%%%%%%%%%%%%%

%%%%%%%%%%%%%%%%%%%%%%%%%%%%%%%%%%%%%%%%%%%%%%%%%%%%%%%%%%%%%%%%%%%%%
\begin{figure}[t!]
\centering
    \includegraphics[width=0.33\linewidth, angle = -90]{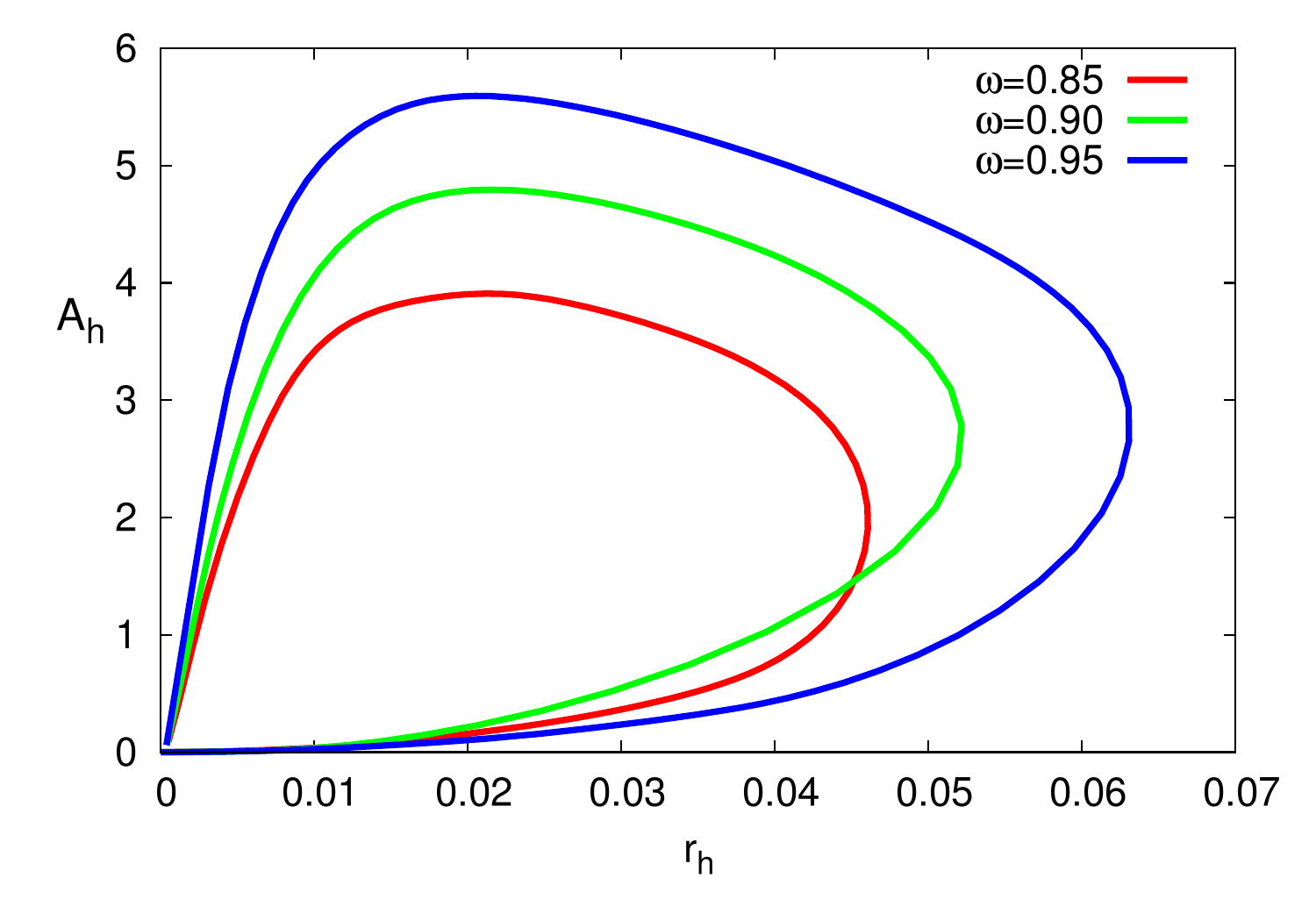}
    \includegraphics[width=0.33\linewidth, angle = -90]{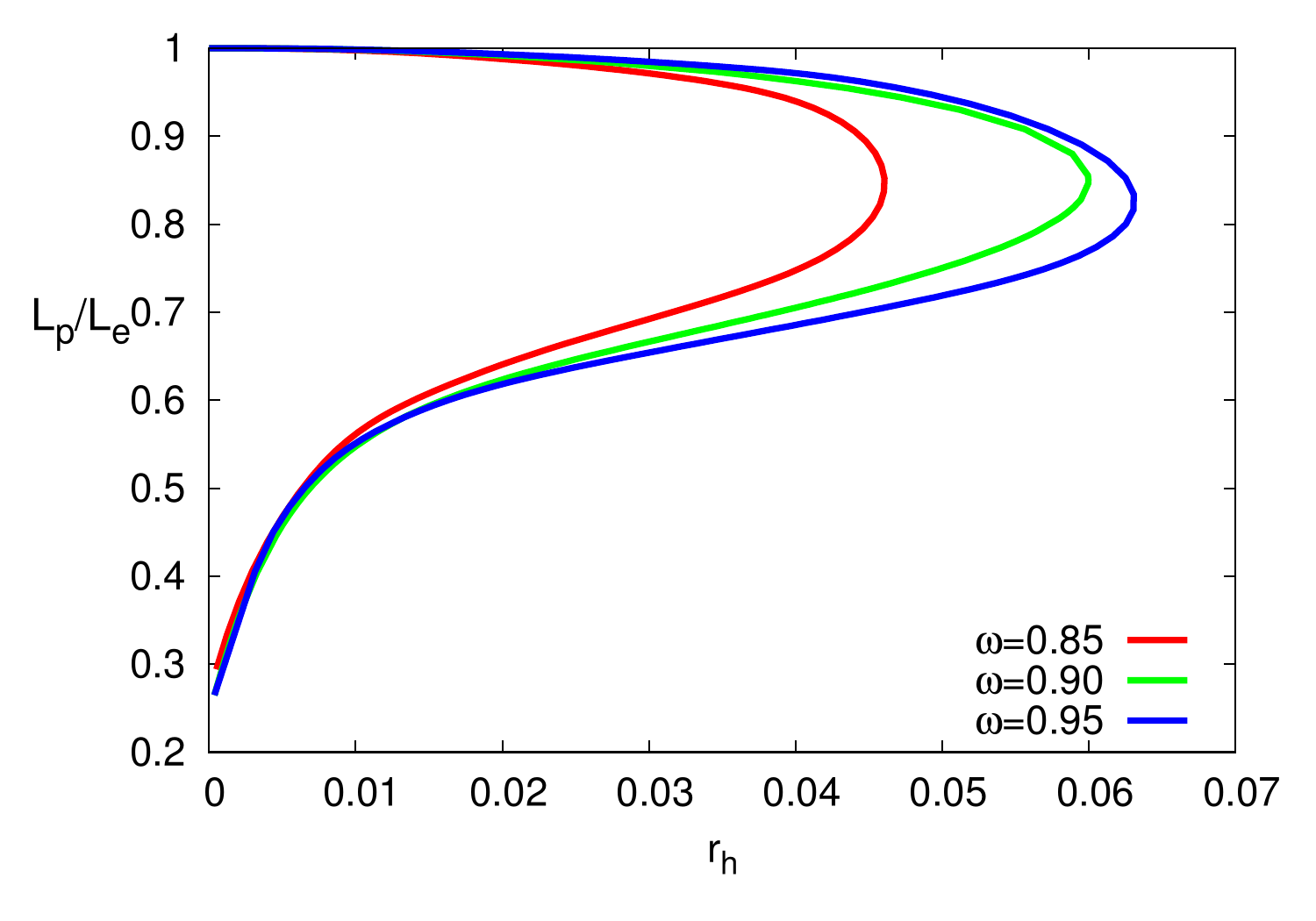}
\caption{The type I $O(3)$ black hole with small event horizon: 
The horizon area $A_h$ (left) and the horizon deformation $\epsilon$ (right) vs the horizon radius $r_h$ for different values of $\omega$, $\beta=0.35$, $\alpha=0.6$.}
    \label{fig_type1_AEvsRh}
\end{figure}\par
%%%%%%%%%%%%%%%%%%%%%%%%%%%%%%%%%%%%%%%%%%%%%%%%%%%%%%%%%%%%%%%%%%%%%

The corresponding horizon area $A_h$ and horizon deformation $\epsilon$ are shown in Fig.~\ref{fig_type1_AEvsRh}.
The dependence of the horizon area on the horizon radius $r_h$ is a loop that shrinks as $\omega$ decreases. 
The horizon deformation increases with $r_h$ along the first branch up to the turning point at $r_h=r_h^{\rm cr}$.
Along the second branch, it then increases as $r_h$ decreases.

\subsection{Black holes with type II hair}

The type II black holes are the black-hole counterparts of the regular scalaron-supported spinning solitons introduced above. 
For these configurations the scalar field takes the vacuum value both on the symmetry axis and at spatial infinity, while in the near-horizon equatorial region it may approach the anti-vacuum. 
Thus the scalar distribution is strongly deformed and remains compatible with the regular $n=1$ axis condition $f=0$.
%%%%%%%%%%%%%%%%%%%%%%%%%%%%%%%%%%%%%%%%%%%%%%%%%%
\begin{figure}[h!]
\begin{center}
\includegraphics[height=.24\textheight,  angle =0]{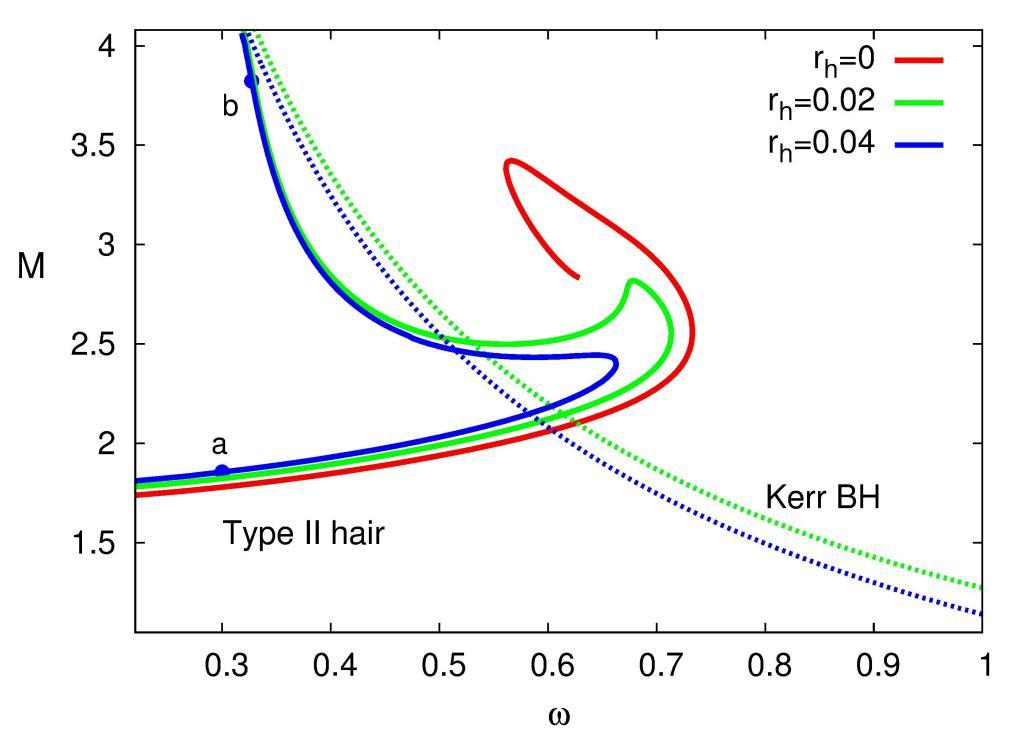}
\includegraphics[height=.24\textheight,  angle =0]{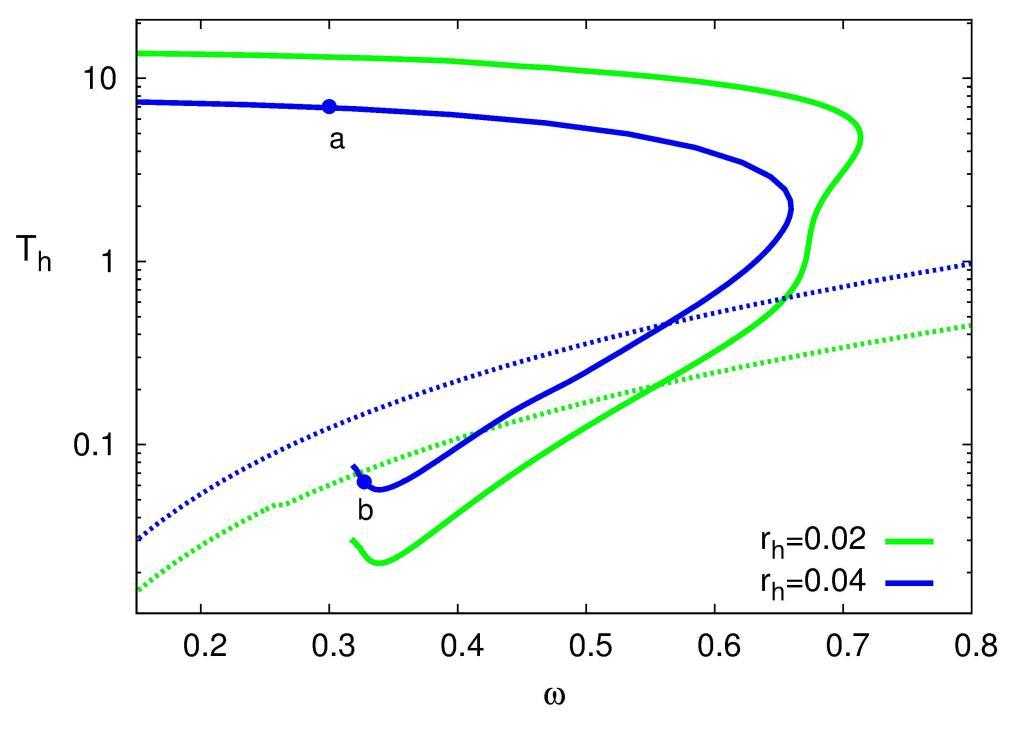}
\end{center}
\caption{\small
Type II $O(3)$ hairy black holes with small event horizon:
ADM mass $M$ in units of $16\pi$ (left plot) and the Hawking temperature in units of $1/16\pi$ (right plot) are displayed vs the angular frequency $\omega$ for $\beta=0.5$ and a set of values of the horizon radius $r_h$ for $\alpha=0.6$.
For comparison the corresponding vacuum Kerr solutions are also shown (dotted lines).}
    \label{figBHII}
\end{figure}
%%%%%%%%%%%%%%%%%%%%%%%%%%%%%%%%%%%%%%%%%%%%%%%%%%%%%%%%%%%%%%%%%%%%%
%%%%%%%%%%%%%%%%%%%%%%%%%%%%%%%%%%%%%%%%%%%%%%%%%%
\begin{figure}[h!]
\begin{center}
\includegraphics[height=.24\textheight,  angle =0]{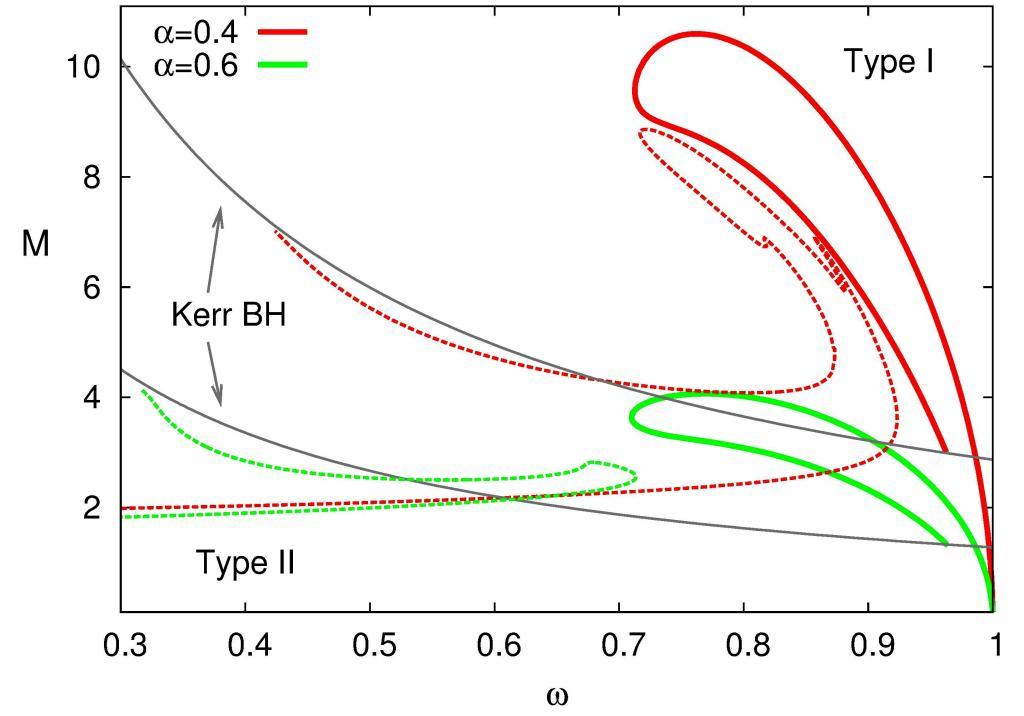}
\includegraphics[height=.24\textheight,  angle =0]{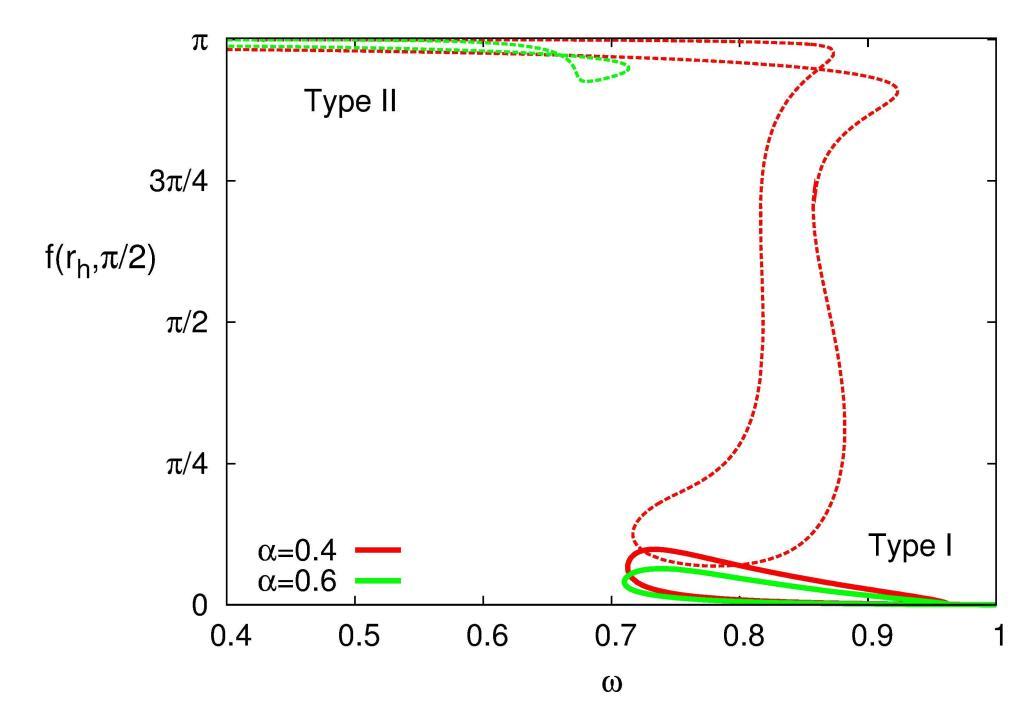}
\end{center}
\caption{\small
Type I and type II $O(3)$ hairy black holes with small event horizon:
ADM mass $M$ in units of $16\pi$ (left plot) and the value of the scalar profile function $f(r_h,\theta=\pi/2)$  (right plot) are displayed vs the angular frequency $\omega$ for $\beta=0.5$ and $r_h=0.02$ for  $\alpha=0.6$ and $\alpha=0.4$.
For comparison the corresponding vacuum Kerr solutions are also shown.}
    \label{figBHIII}
\end{figure}
%%%%%%%%%%%%%%%%%%%%%%%%%%%%%%%%%%%%%%%%%%%%%%%%%%%%%%%%%%%%%%%%%%%%%

%%%%%%%%%%%%%%%%%%%%%%%%%%%%%%%%%%%%%%%
\begin{figure}[t!]
\begin{center}
\includegraphics[height=.20\textheight,  angle =-0]{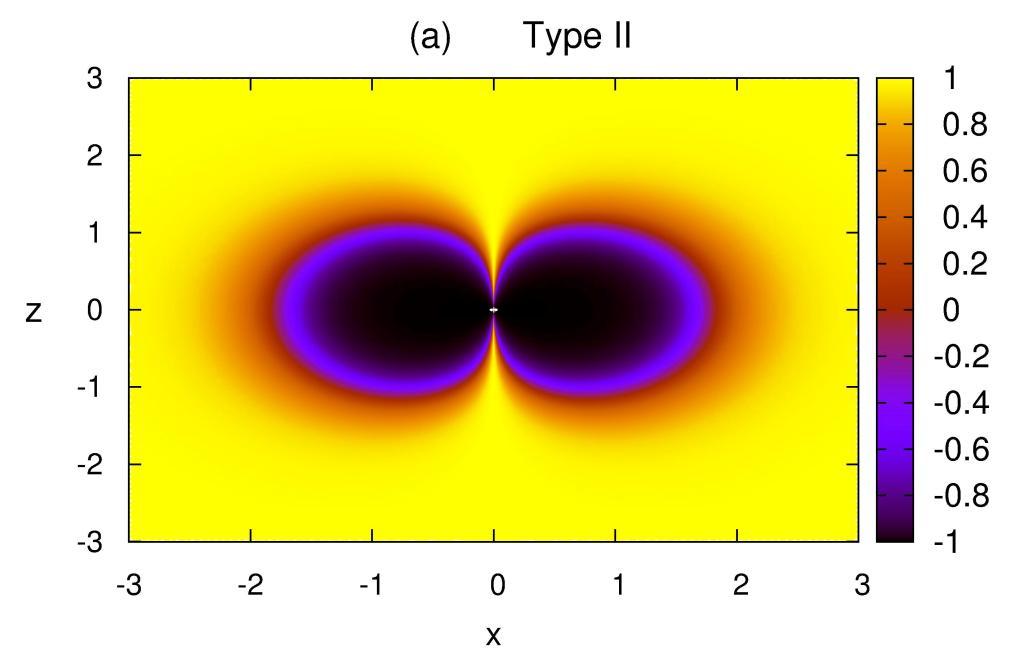}
\includegraphics[height=.20\textheight,  angle =-0]{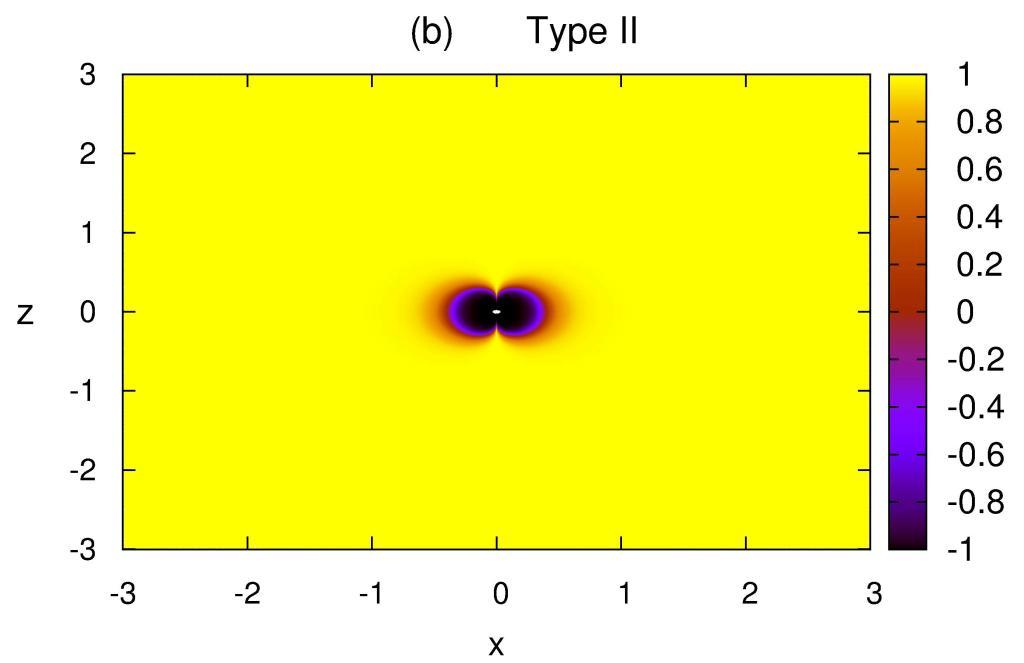}
\end{center}
\caption{\small
The type II $O(3)$ hairy black holes with small event horizon:
Contour plots of the scalar field component $\phi_3$ in the $x z$ plane are displayed for the configurations, indicated by the corresponding dots in the left panel of Fig.~\ref{figBHII}, for $\beta=0.5$ and the horizon radius $r_h=0.04$. 
}
    \label{fig_contour_2}
\end{figure}
%%%%%%%%%%%%%%%%%%%%%%%%%%%%%%%%%%%%%%%%%%%%%%%%%%%%%%%%%%%%%%%%%%%%%

%%%%%%%%%%%%%%%%%%%%%%%%%%%%%%%%%%%%%%%%%%%%%%%%%%%%%%%%%%%%%%%%%%%%%%
\begin{figure}[t!]
\begin{center}
\includegraphics[height=.34\textheight,  angle =-90]{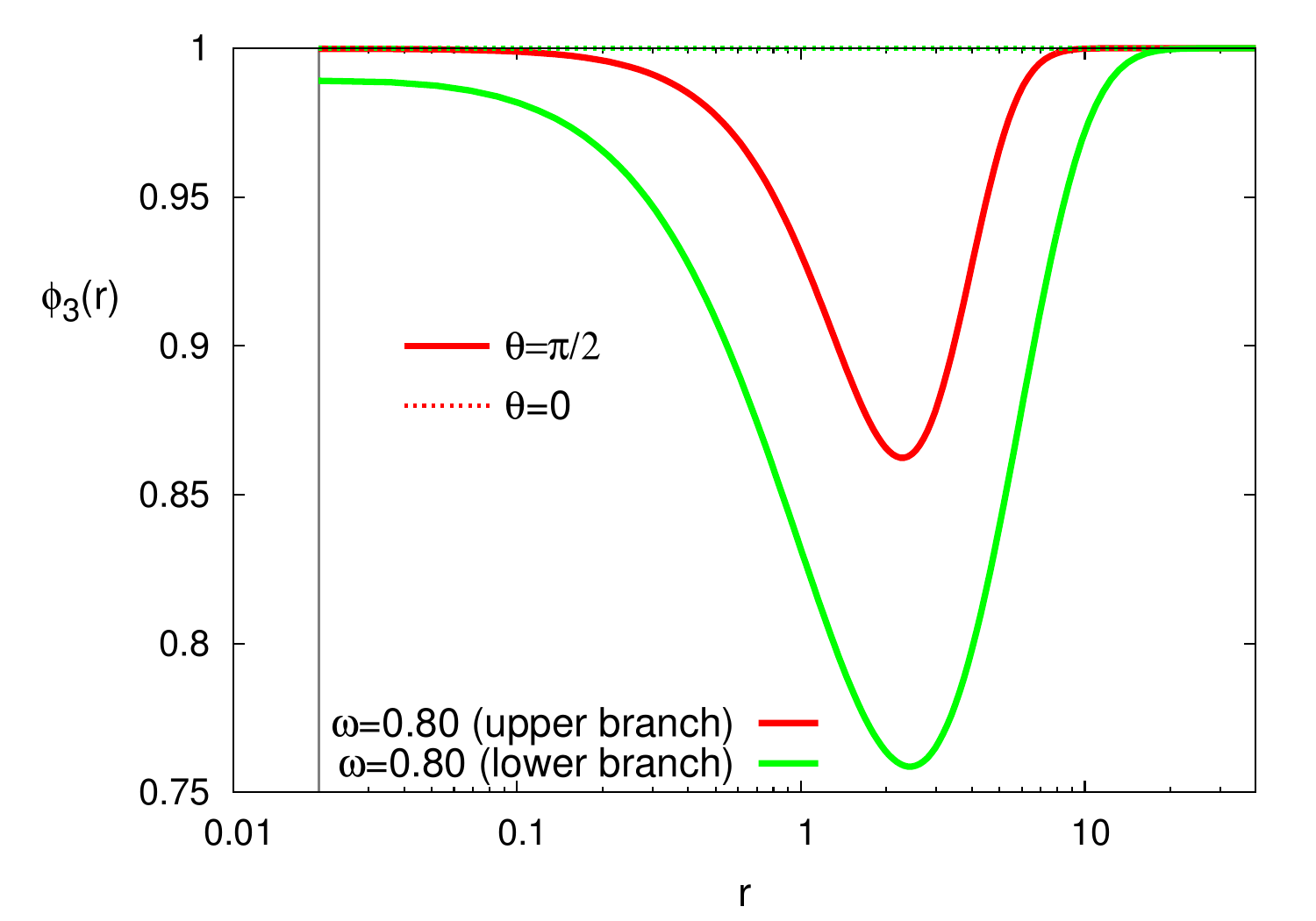}
\includegraphics[height=.34\textheight,  angle =-90]{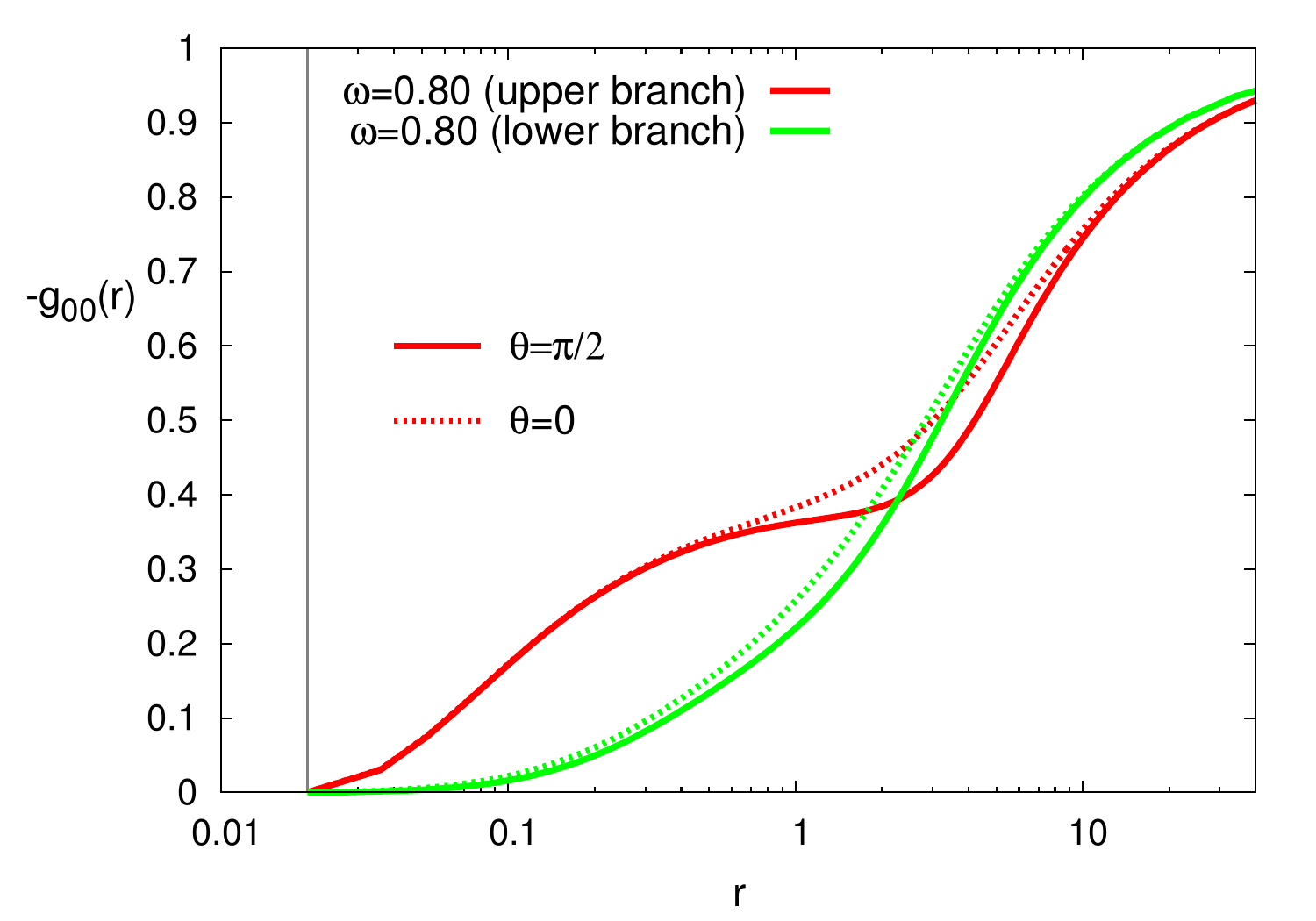}
\includegraphics[height=.34\textheight,  angle =-90]{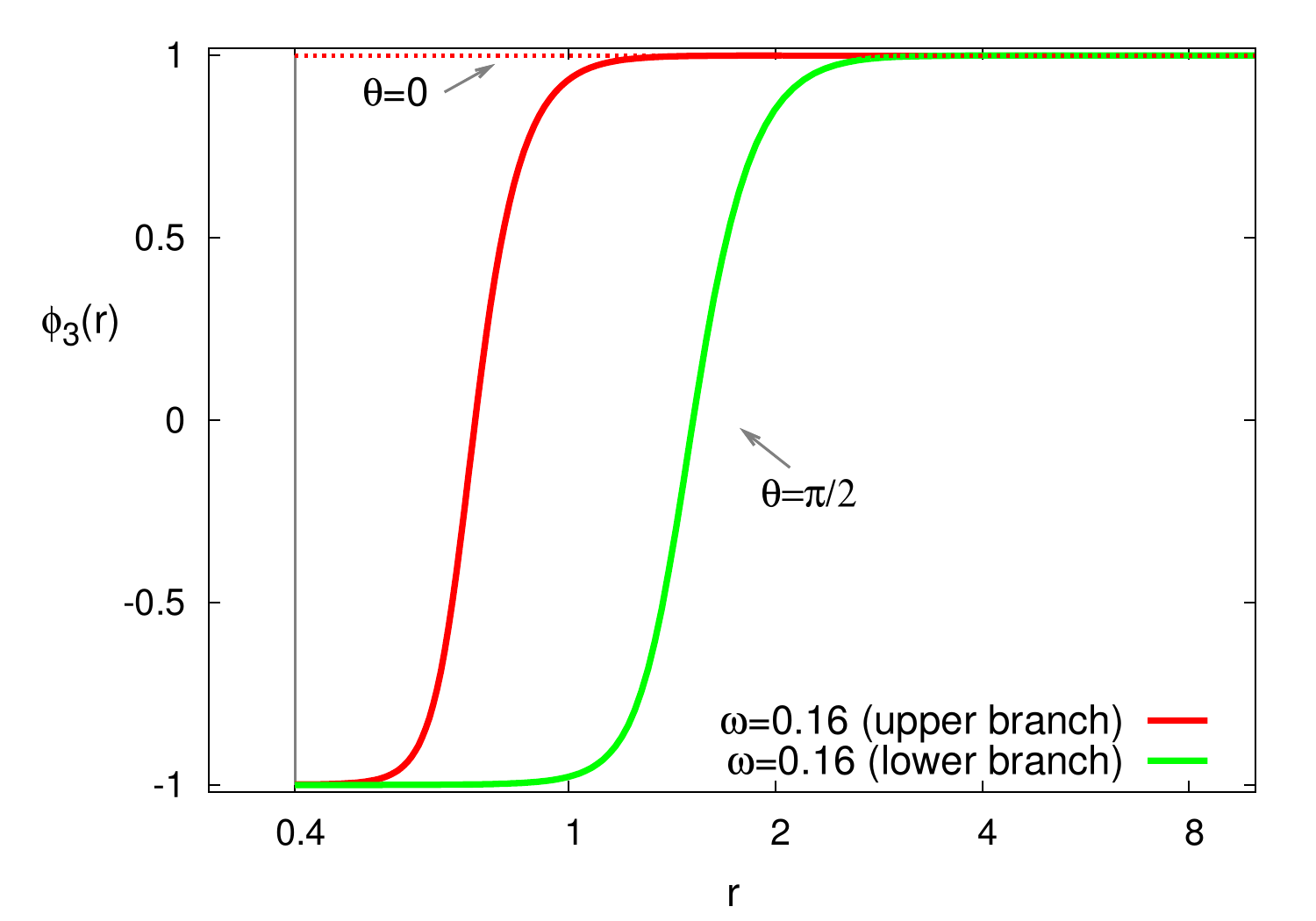}
\includegraphics[height=.34\textheight,  angle =-90]{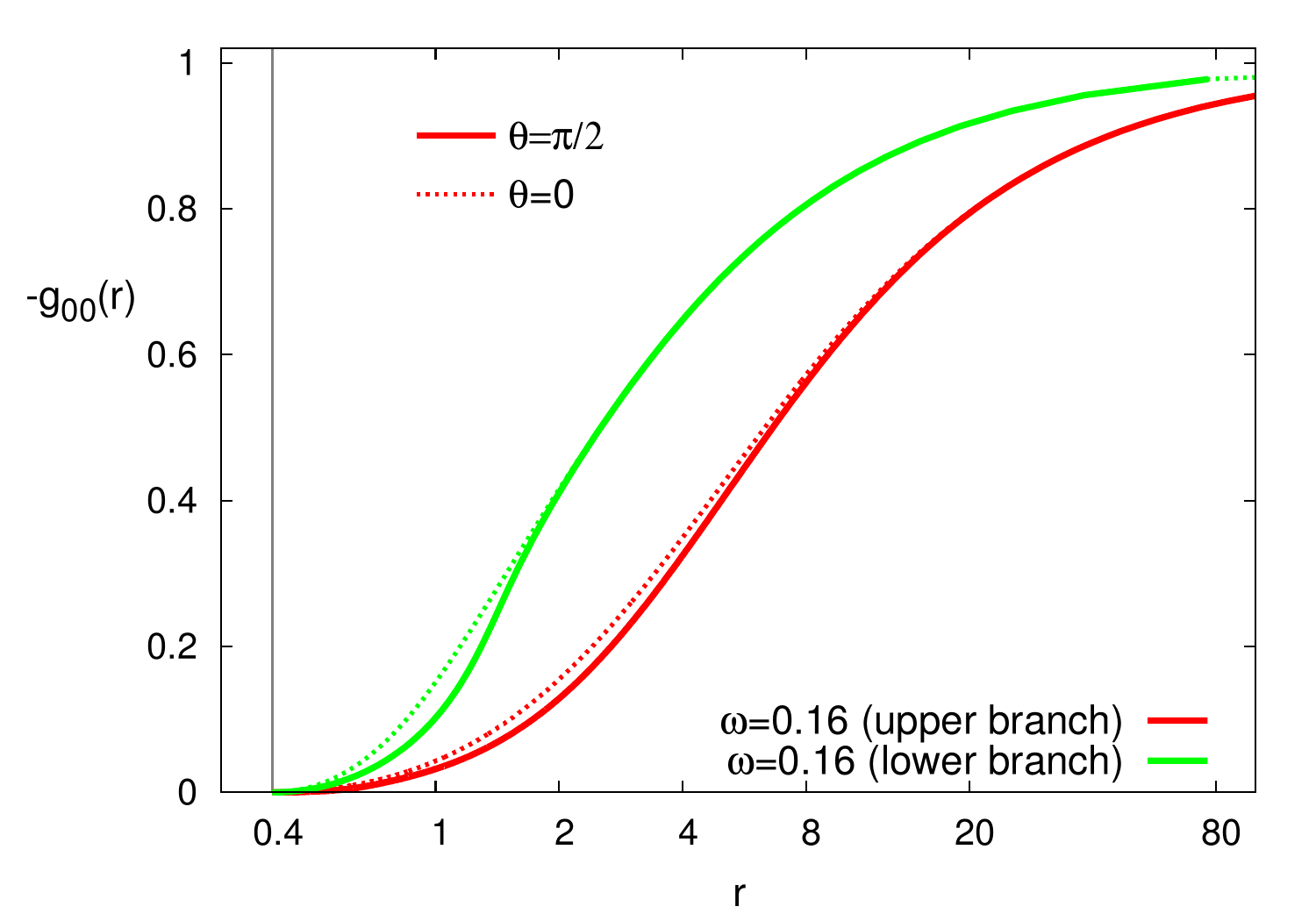}
\end{center}
\caption{\small
Two types of $O(3)$ black holes with hair. 
Left column: profiles of the scalar field component $\phi_3$ for $\beta=0.5$ and $\alpha=0.6$ on the upper and lower branches at $\omega=0.80$ (Type I), $r_h=0.02$, and at $\omega=0.16$ for $r_h=0.4$ (Type II). 
Right column: the radial profiles of the metric function $-g_{00}(r)$ for the corresponding configurations.
Solid lines correspond to the value of polar angle $\theta =\pi/2$, dotted lines correspond to $\theta=0$. 
}
    \label{fig_comp}
\end{figure}
%%%%%%%%%%%%%%%%%%%%%%%%%%%%%%%%%%%%%%%%%%%%%%%%%%%%%%%%%%%%%%%%%%%%%

Figure \ref{figBHII} shows the behavior of $O(3)$-symmetric hairy black holes of type II with a small event horizon (compare with the corresponding plots for type I, Fig.~\ref{figBH1}).
The type II black holes with small horizon radii exhibit a particularly interesting behavior as the frequency decreases. 
For $r_h=0.02$, the ADM mass initially increases steeply with decreasing $\omega$, but around $\omega\simeq 0.3$ the mass-frequency curve changes direction rather sharply and subsequently runs approximately parallel to the corresponding Kerr curve. 
The crossing of the Kerr mass curve does not signal a transition to a vacuum black hole, since the scalar field remains nontrivial, as is also evident from the equatorial horizon profile shown in Fig.~~\ref{figBHIII}.
Moreover, the Hawking temperature exhibits a markedly different behavior from that of the corresponding Kerr solutions. 
The scalar field distributions in Fig.~\ref{fig_contour_2} illustrate that, along the type II family, the scalar configuration can become strongly localized towards the horizon. 
This suggests that the pronounced change in the mass-frequency curve is associated with a rapid localization of the scalar hair rather than its disappearance. 
Our numerical solutions at small but finite frequencies show that, as $\omega$ decreases, the scalar field remains close to its anti-vacuum value $f=\pi$ over much of the horizon, while the region where it approaches its vacuum value $f=0$ extends progressively closer to the horizon. 
Thus, the nontrivial scalar configuration becomes increasingly confined to a narrow region near the horizon, although its precise behavior in the limit $\omega\to 0$ remains undetermined.
The global properties of these configurations may become approximately Kerr-like while significant differences persist in the near-horizon region.

The qualitative difference between the two regular families is
illustrated in Figs.~\ref{figBHIII} and \ref{fig_comp}. 
For the parameter values considered in Fig.~\ref{figBHIII}, the type I and type II black hole configurations remain distinguishable by their scalar field profiles. 

For smaller values of the gravitational coupling, the branch structure of the type II black holes appears to become more complicated. 
In particular, for $\alpha=0.4$ and $\beta=0.5$, the numerical results suggest the presence of additional branches, accompanied by pronounced changes in the scalar profile function at the equatorial horizon, $f(r_h,\pi/2)$, as illustrated in Fig.~\ref{figBHIII}.
Along the backward branches, some type II black hole solutions approach the corresponding Kerr curves in the mass-frequency diagram.
However, this proximity in the global charges does not necessarily imply that the solutions converge to Kerr black holes. 
The scalar field may remain nontrivial but become strongly localized near the horizon, as illustrated by the representative configurations in Fig.~\ref{fig_contour_2}.
On the other hand, the type I configurations are trivializing on the Kerr line.

The upper panels in Fig.~\ref{fig_comp} show representative type I solutions on the two branches. 
Their scalar field remains relatively close to the vacuum configuration. 
In contrast, the type II solutions shown in the lower panels develop a pronounced near-horizon equatorial region in which $\phi_3$ approaches the anti-vacuum value $\phi_3=-1$. 
The corresponding metric profiles also differ significantly for the two families.

\begin{figure}[h!]
\begin{center}
\includegraphics[height=.24\textheight,  angle =0]{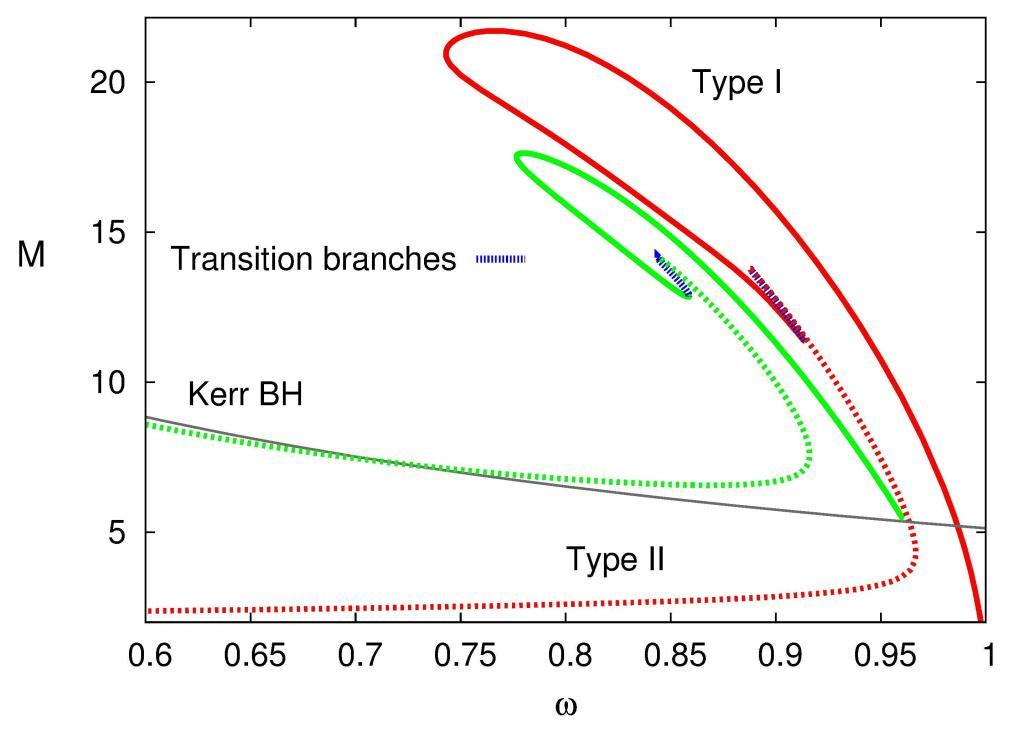}
\includegraphics[height=.24\textheight,  angle =0]{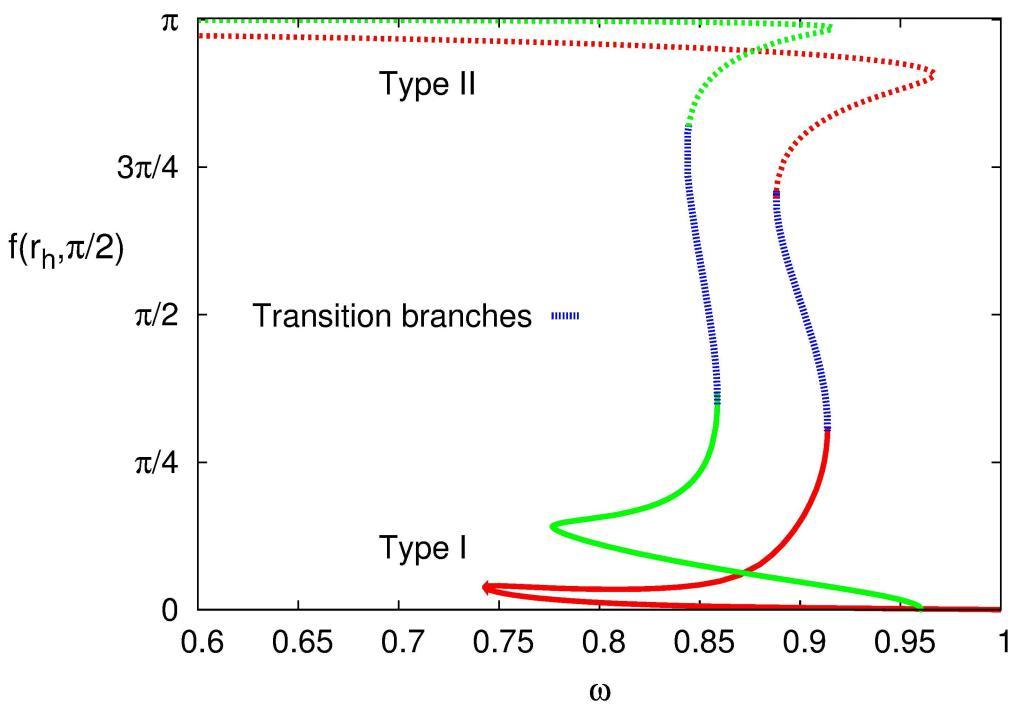}
\includegraphics[height=.24\textheight,  angle =0]{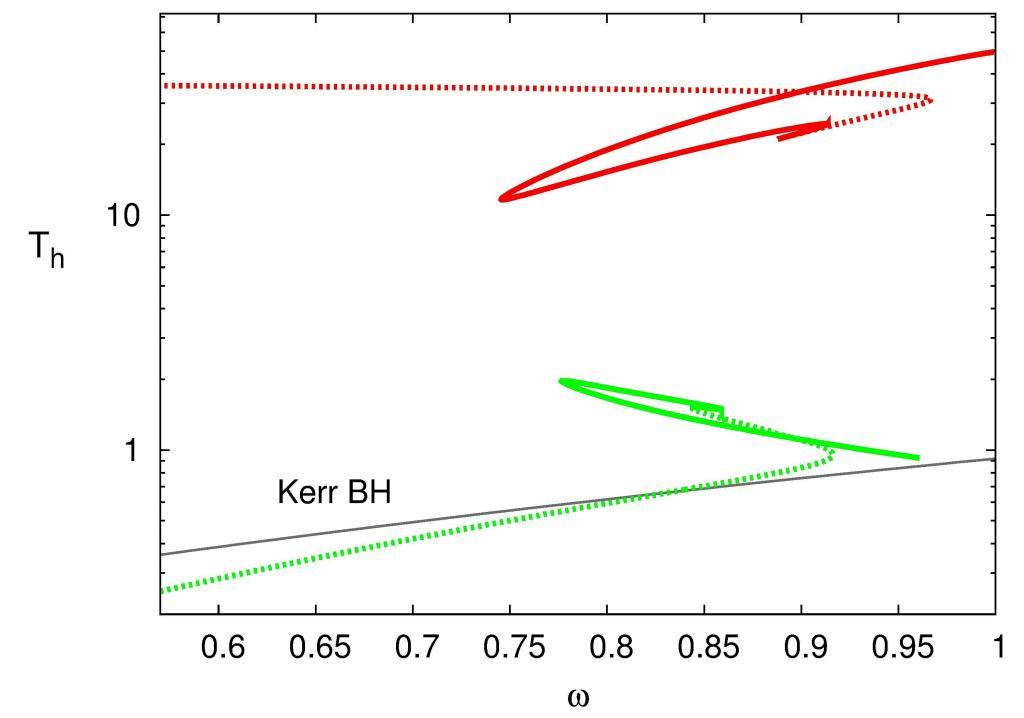}
\end{center}
\caption{\small
%Bifurcating 
Reconnected branches of type I and type II $O(3)$ hairy black holes with small event horizon:
ADM mass $M$ in units of $16\pi$ (upper left plot) and the value of the scalar profile function $f(r_h,\theta=\pi/2)$  (upper right plot) and the Hawking temperature in units of $1/16\pi$ (bottom plot)  are displayed vs the angular frequency $\omega$ for $\beta=0.5$ and $r_h=0.02$ for  $\alpha=0.3$.
For comparison the corresponding vacuum Kerr solutions are also shown.
The red curves represent the family connecting the fundamental type I and type II branches, while the green curves represent the family connecting the corresponding second branches. 
The short blue segments indicate the transition regions between type I and type II configurations.
}
    \label{figBHIV}
\end{figure}

By analogy with the regular solutions discussed above, the type I and type II black-hole branches undergo a merger and reconnection when the effective gravitational coupling decreases below a critical value, which we estimate as $\alpha_{\rm cr}\simeq 0.3707$ for $\beta=0.5$ and $r_h=0.02$.
As the gravitational coupling decreases through the critical value, the original connections between the fundamental and second branches within each type are lost. 
Instead, the fundamental branches of type I and type II reconnect to form one continuous family, while the corresponding second branches reconnect to form another continuous family. 
This change in the branch connectivity is illustrated in Fig.~\ref{figBHIV} for $\alpha=0.3$. 
The continuous sequence of corresponding field configurations begins as perturbative excitations above the vacuum when the frequency $\omega$ drops below the mass threshold, and, following a series of deformations, transitions into the second type of solutions, which exist up to the static limit $\omega \to 0$. 
At the same time, the second connected family of solutions emerges as a perturbative branch from the Kerr solutions at a critical frequency $\omega_{\mathrm{cr}}<1$. 
Further variation of the frequency $\omega$ reveals a sequence of interconnected solution branches within this family, which evolve toward a limiting configuration of a scalar field that is essentially localized at the event horizon in the anti-vacuum state, see Fig.~\ref{figBHIV}.
The two families formed by the reconnection also differ in their Hawking temperatures, as shown in the lower panel of Fig.~\ref{figBHIV}. 
At comparable frequencies, the family connecting the fundamental type I and type II branches has significantly higher temperatures than the family connecting the corresponding second branches.

%%%%%%%%%%%%%%%%%%%%%%%%%%%%%%%%%%%%%%%%%%%%%%%%%%
%%%%%%%%%%%%%%%%%%%%%%%%%%%%%%%%%%%%

%%%%%%%%%%%%%%%%%%%%%%%%%%%%%%%%%%%%%%%%%%%%%%%%%%%%%%%%%%%%%%%%%%%%%%
\begin{figure}[t!]
\begin{center}
\includegraphics[height=.34\textheight,  angle =-90]{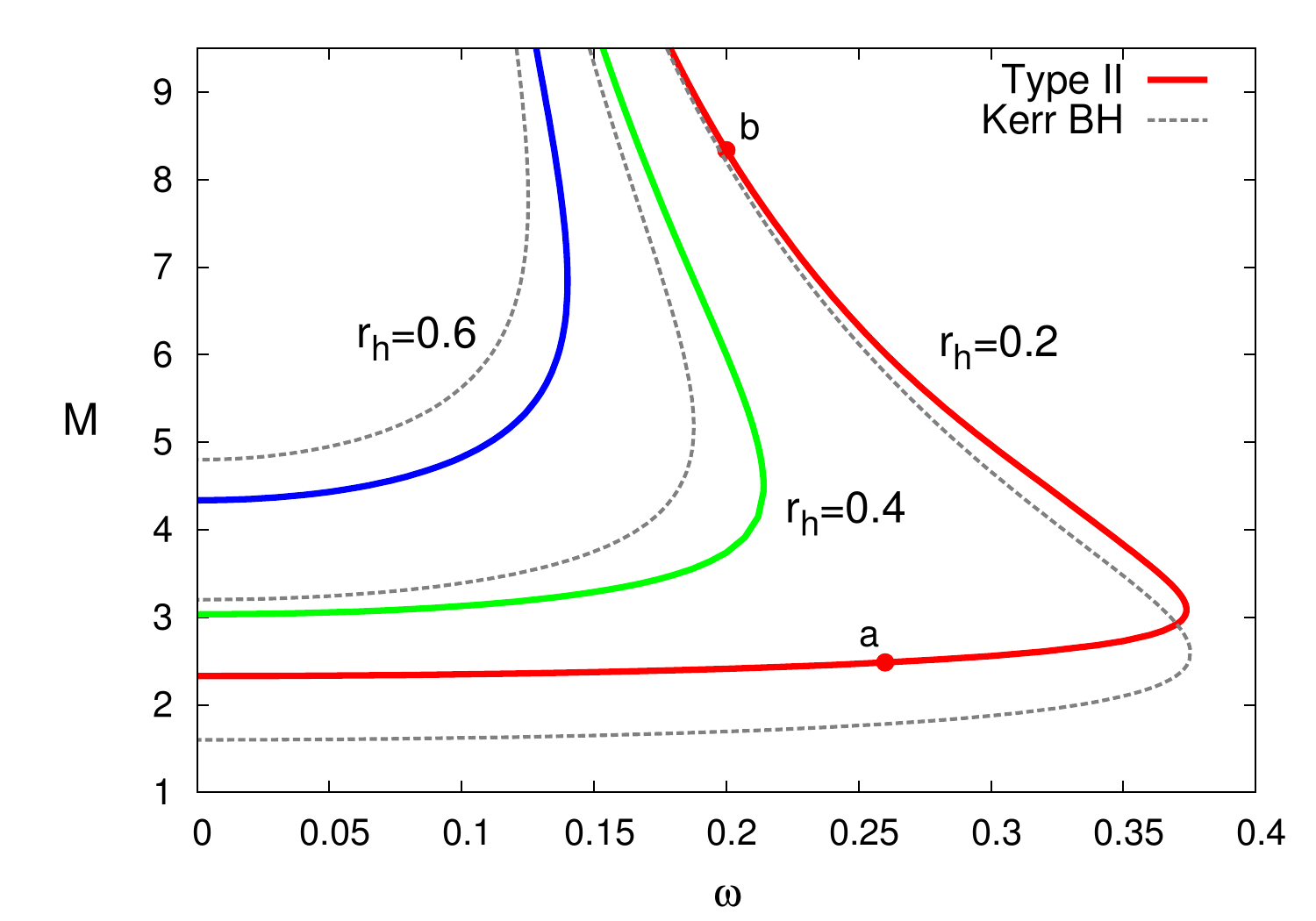}
\includegraphics[height=.34\textheight,  angle =-90]{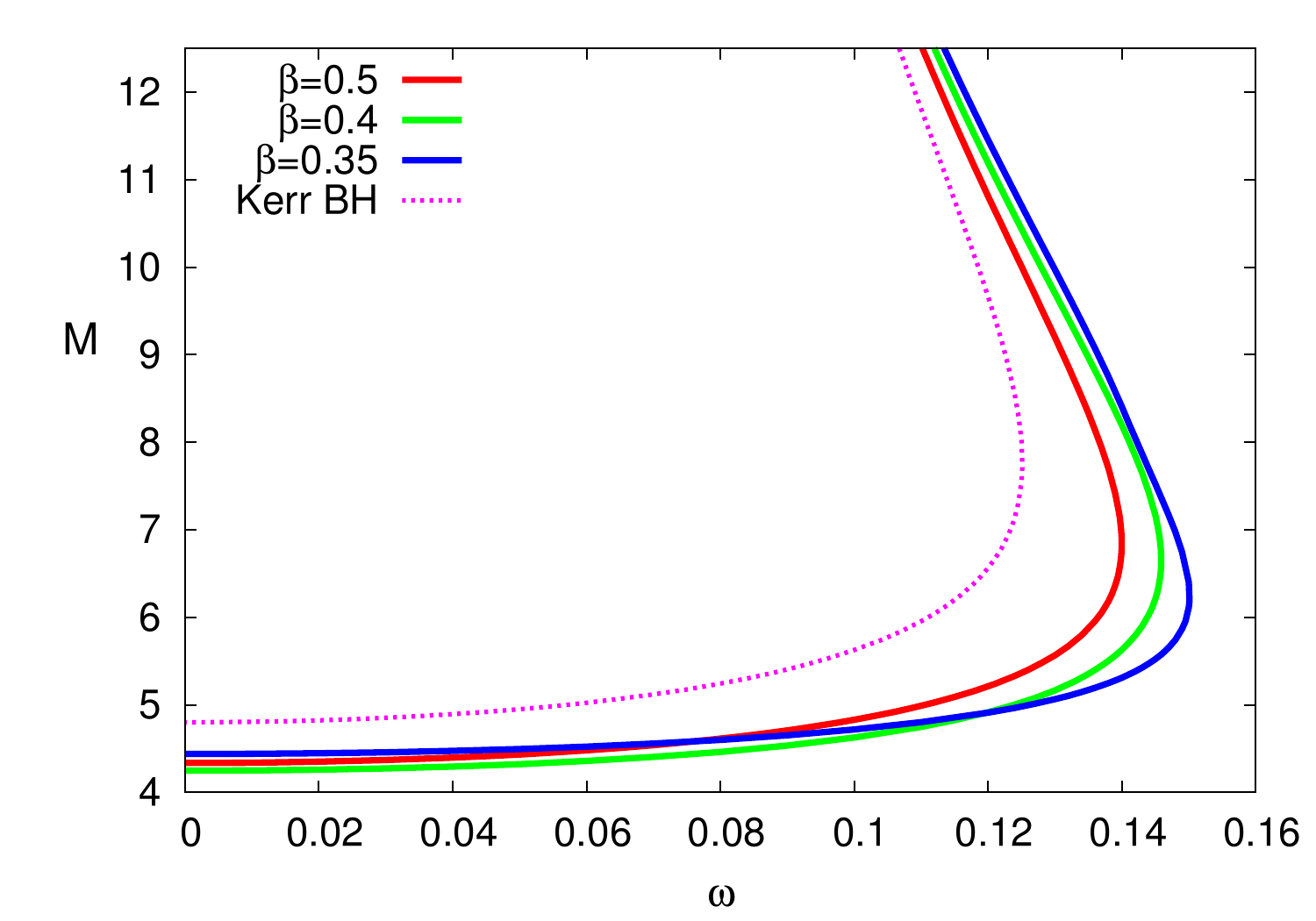}
\end{center}
\caption{\small
Type II $O(3)$ hairy black holes with large event horizon:
ADM mass $M$ in units of $16\pi$ is displayed vs the angular
frequency $\omega$ for $\alpha=0.5$, $\beta=0.5$ and a set of values of the horizon radius $r_h$ (left), and for a set of values of the parameter $\beta$ at $\alpha=0.5$ and $r_h=0.6$ (right).
For comparison the corresponding Kerr solutions are also shown (dotted lines).
}
    \label{fig3}
\end{figure}
%%%%%%%%%%%%%%%%%%%%%%%%%%%%%%%%%%%%%%%%%%%%%%%%%%%%%%%%%%%%%%%%%%%%%

The regular type II black holes may extend to substantially larger horizon radii than the type I configurations.
Figure~\ref{fig3} displays the ADM mass of these large-horizon type II solutions versus the angular frequency.

%%%%%%%%%%%%%%%%%%%%%%%%%%%%%%%%%%%%%%%%%%%%%%%%%%%%%%%%%%%%%%%%%%%%%%
\begin{figure}[t!]
\begin{center}
\includegraphics[height=.24\textheight,  angle =0]{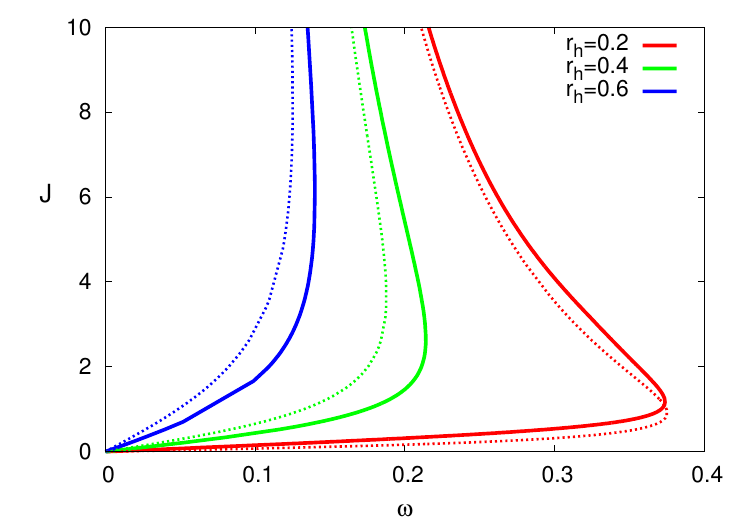}
\includegraphics[height=.24\textheight,  angle =0]{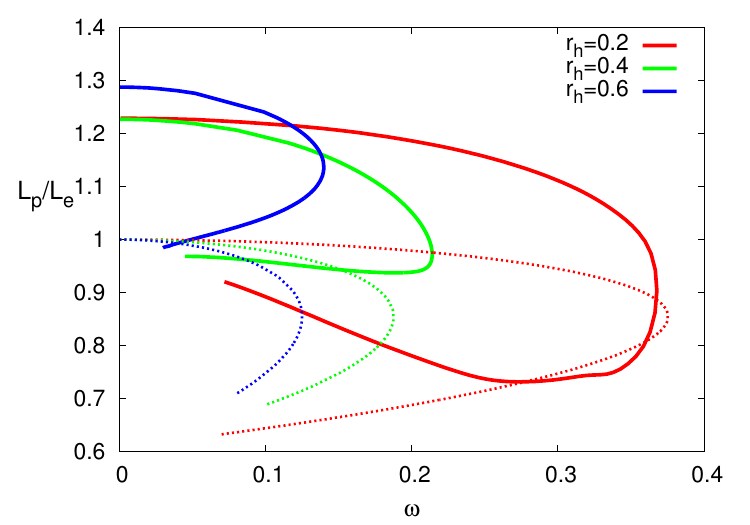}
\includegraphics[height=.24\textheight,  angle =0]{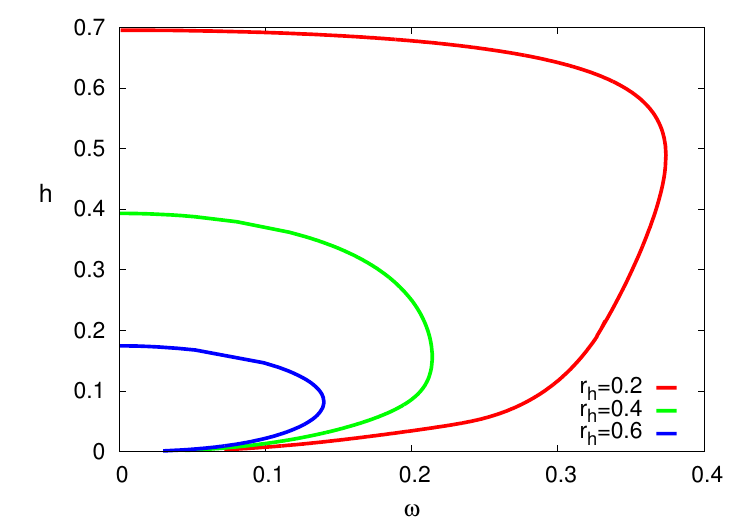}
\end{center}
\caption{\small
Type II $O(3)$ hairy black holes with large event horizon:
Angular momentum $J$ in units of $16\pi$ (left plot), ratio of polar to equatorial circumferences (right plot) and ``hairiness'' $h$ (bottom plot) are displayed vs the angular frequency $\omega$ for a set of values of the horizon radius $r_h$ at $\beta=0.5$, $\alpha=0.5$.
For comparison the corresponding Kerr solutions are also shown (dotted lines).
}
    \label{fig_JvsW_typeII}
\end{figure}
%%%%%%%%%%%%%%%%%%%%%%%%%%%%%%%%%%%%%%%%%%%%%%%%%%%%%%%%%%%%%%%%%%%%%

%%%%%%%%%%%%%%%%%%%%%%%%%%%%%%%%%%%%%%%%%%%%%%%%%%%%%%%%%%%%%%%%%%%%%%
\begin{figure}[t!]
\begin{center}
\includegraphics[height=.22\textheight,  angle =-0]{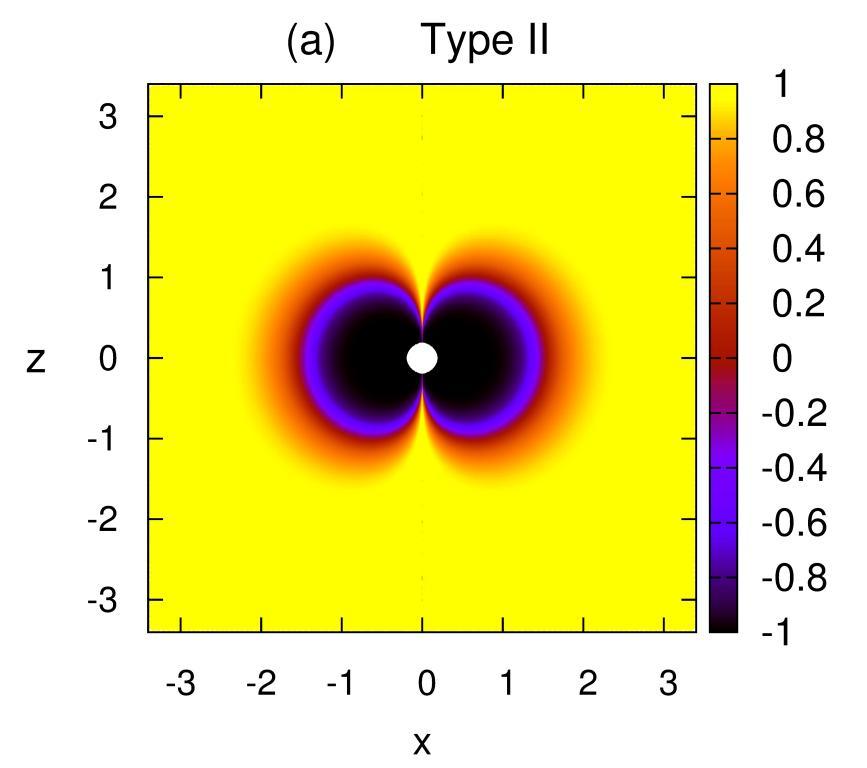}
\includegraphics[height=.22\textheight,  angle =-0]{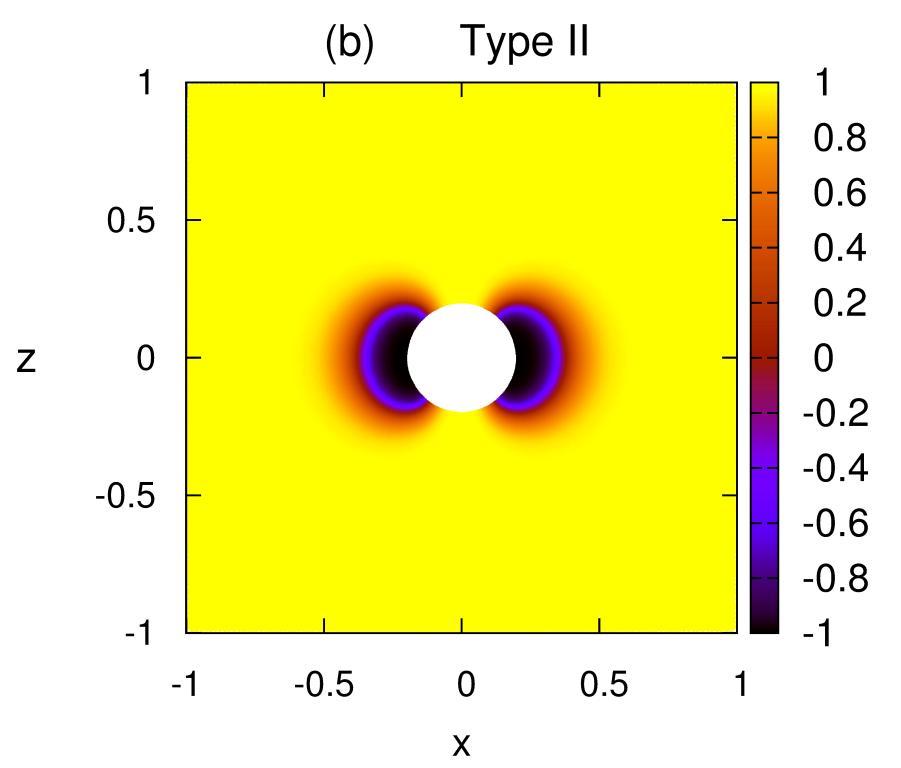}
\end{center}
\caption{\small
Type II $O(3)$ hairy black holes with large event horizon:
Contour plots of the scalar field component $\phi_3$ for configurations indicated by the corresponding dots in Fig.~\ref{fig3}, for $\beta=0.5$ and the horizon radius $r_h=0.2$.
}
    \label{fig4}
\end{figure}
%%%%%%%%%%%%%%%%%%%%%%%%%%%%%%%%%%%%%%%%%%%%%%%%%%%%%%%%%%%%%%%%%%%%%

As the horizon radius increases, the type II scalar distribution becomes strongly deformed. The profile function $f$ vanishes along the symmetry axis and, in the equatorial plane, interpolates between the anti-vacuum in the near-horizon region and the vacuum at spatial infinity; see Fig.~\ref{fig_comp}.

%%%%%%%%%%%%%%%%%%%%%%%%%%%%%%%%%%%%%%%%%%%%%%%%%%%%%%%%%%%%%%%%%%%%%%
\begin{figure}[t!]
\begin{center}
\includegraphics[height=.34\textheight,  angle =-90]{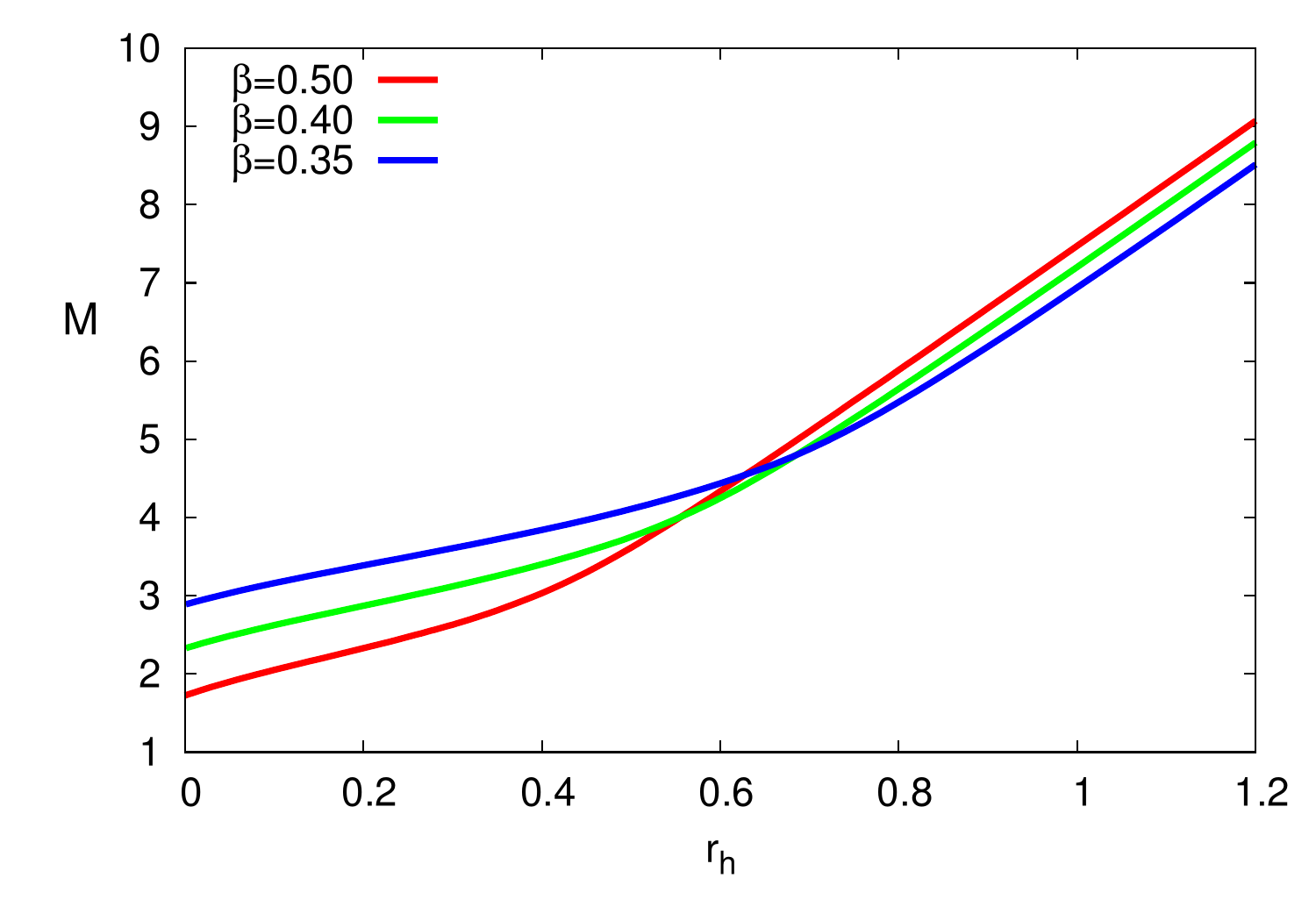}
\includegraphics[height=.34\textheight,  angle =-90]{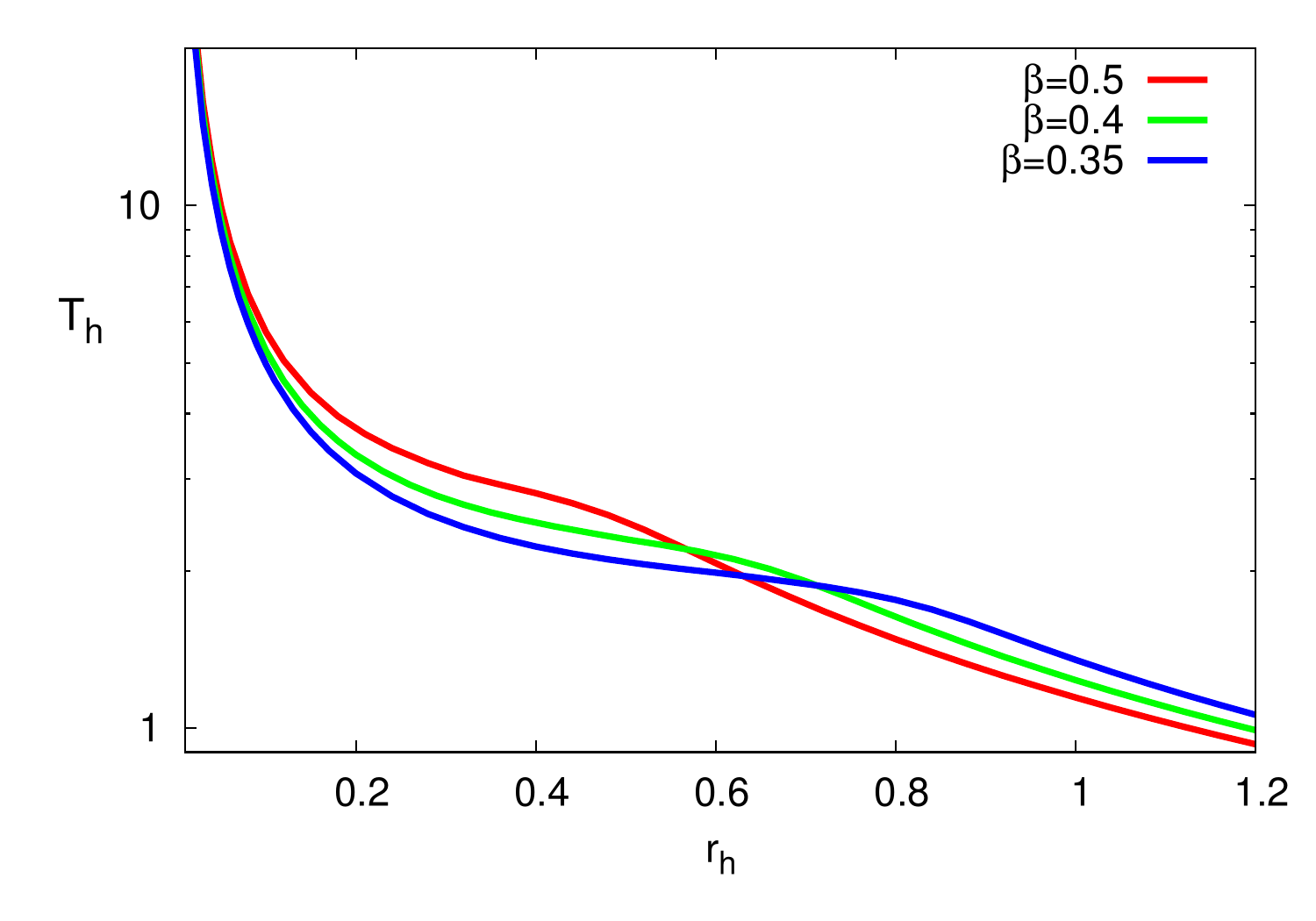}
\end{center}
\caption{\small
Type II $O(3)$ hairy black holes in the non-rotating limit $\omega=0$: the ADM mass $M$ in units of $16\pi$ (left) and Hawking temperature $T_h$ in units of $1/16\pi$  (right) versus the horizon radius $r_h$ for $\alpha=0.5$ and several values of $\beta$. 
}
    \label{fig3a}
\end{figure}
%%%%%%%%%%%%%%%%%%%%%%%%%%%%%%%%%%%%%%%%%%%%%%%%%%%%%%%%%%%%%%%%%%%%%

Solutions of this type exist for sufficiently large values of the effective gravitational coupling $\alpha$. 
In general there are two $\omega$-branches, which bifurcate at a maximal frequency $\omega_{\max}^{II}$. 
The maximal frequency increases as the parameter $\beta$ decreases, and the scalar hair becomes smaller as the solutions approach the corresponding Kerr configurations.
The corresponding angular momentum, horizon deformation, and hairiness are displayed in Fig.~\ref{fig_JvsW_typeII}. 
The two-branch structure is clearly visible also in these quantities.
The horizon deformation can be either prolate or oblate, depending on the position along the branches.

The behavior of the upper Kerr curves shown in Fig.~\ref{fig3} for comparison can be understood as follows. 
In the isotropic coordinates employed, the Kerr solutions satisfy
\be
2r_h=\sqrt{(GM)^2-a^2},
\qquad
\Omega_h=\frac{a}{2GM(GM+2r_h)} .
\ee
Eliminating $a$ and identifying $\omega=\Omega_h$ yields
\be
\omega=
\frac{\sqrt{(GM)^2-4r_h^2}}
     {2GM(GM+2r_h)} .
\ee
On the upper Kerr branch therefore
\be
\frac{M}{16\pi}\simeq
\frac{1}{2\alpha^2\omega},
\qquad \omega\to0 .
\ee
Thus, at fixed $r_h$, the upper Kerr branch extends towards arbitrarily large mass as $\omega \rightarrow 0$.
The scalar field distributions of two representative type II black holes, indicated by the points in Fig.~\ref{fig3}, are shown in Fig.~\ref{fig4}. 
They illustrate the pronounced difference in the spatial extent of the scalar hair between these configurations.

Static non-rotating ($\omega=0$) type II black holes may also exist when the gravitational coupling is sufficiently large. 
As shown in Fig.~\ref{fig3a}, the type II horizon radius can become very large, with the ADM mass increasing and the Hawking temperature decreasing as it grows.

In the small-horizon limit the type II solutions approach the regular scalaron-supported spinning solitonic branch discussed above.

%%%%%%%%%%%%%%%%%%%%%%%%%%%%%%%%%%%%%%%%%%%%%%%%%%%%%%%%%%%%%%%%%%%%%%
\begin{figure}[t!]
\begin{center}
\includegraphics[height=.34\textheight,  angle =-90]{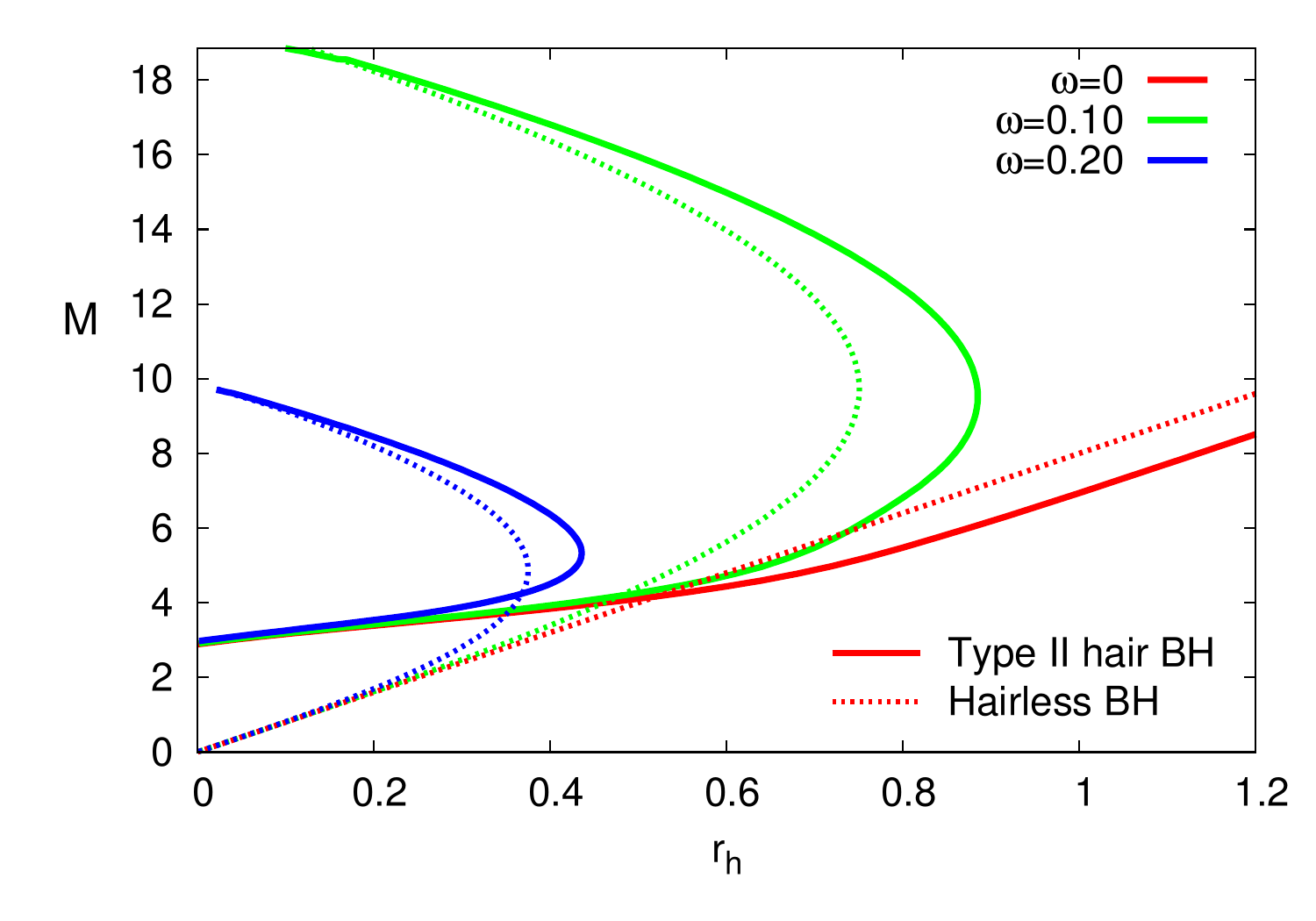}
\includegraphics[height=.34\textheight,  angle =-90]{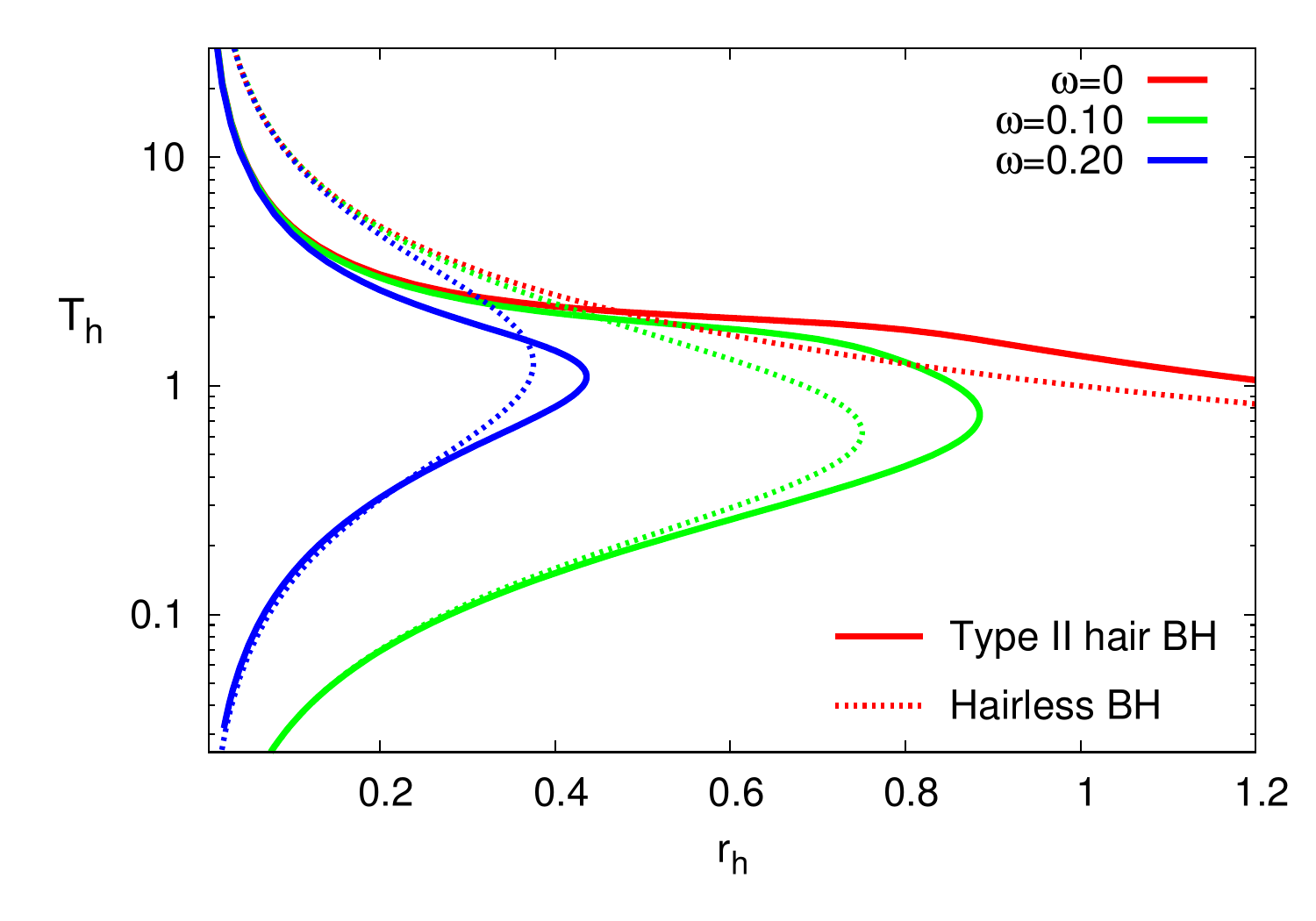}
\end{center}
\caption{\small
Type II $O(3)$ hairy black holes: the ADM mass $M$ in units of $16\pi$  (left) and Hawking temperature $T_h$ in units of $1/16\pi$  (right) versus the horizon radius $r_h$ for $\alpha=0.5$, $\beta=0.35$ and several values of the frequency $\omega$. 
}
    \label{fig3b}
\end{figure}
%%%%%%%%%%%%%%%%%%%%%%%%%%%%%%%%%%%%%%%%%%%%%%%%%%%%%%%%%%%%%%%%%%%%%
 
Figure~\ref{fig3b} shows the ADM mass and Hawking temperature as functions of the horizon radius for several frequencies.
The minimum mass increases with $\omega$, while the usual two-branch pattern is recovered.
For $\omega=0$, the hairless reference solutions are Schwarzschild black holes in isotropic coordinates.
Along this hairless branch the ADM mass grows linearly with the horizon radius, while the Hawking temperature is inversely proportional to the isotropic-coordinate horizon radius $r_h$, i.e., in the normalization used in Fig.~\ref{fig3b},
\be
\frac{M}{16\pi}=\frac{2r_h}{\alpha^2},
\qquad
16\pi T_h=\frac{1}{r_h}.
\ee
The Schwarzschild branch originates from the flat space limit at $r_h\to 0$, where $M\to 0$ and $T_h \to \infty$.
For nonzero values of the frequency $\omega$, the corresponding hairless reference solutions are Kerr black holes. 
The type II hairy black-hole solutions form branches which extend from the corresponding regular type II solitons in the limit $r_h\to 0$ up to a maximal horizon radius.
In the $M(r_h)$ diagram these branches cross the corresponding hairless Kerr branches, while the hairy and hairless solutions remain distinct, see Fig.~\ref{fig3b}.

\begin{figure}[t!]
    \centering
    \includegraphics[width=0.33\linewidth, angle = -90]{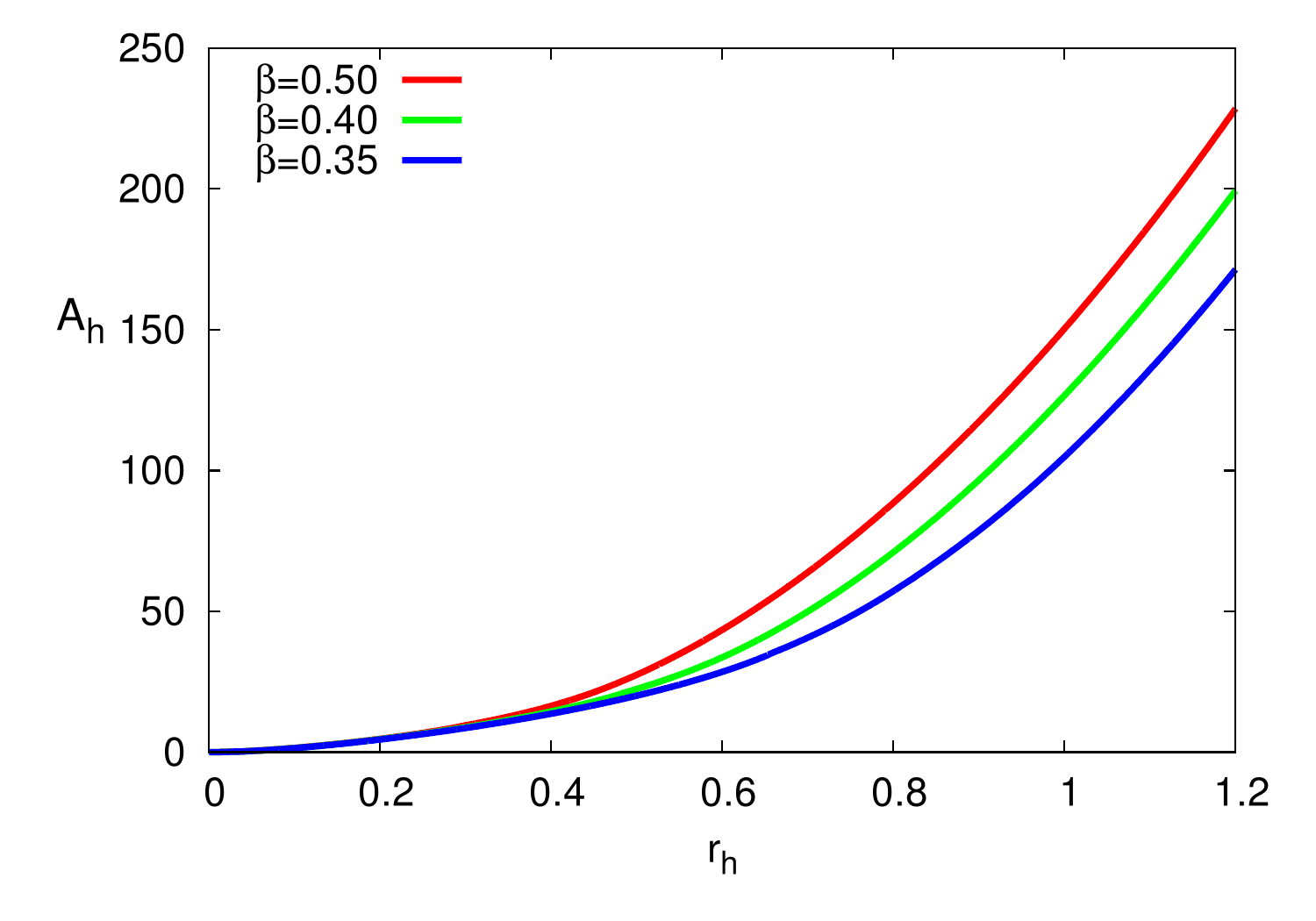}
    \includegraphics[width=0.33\linewidth, angle = -90]{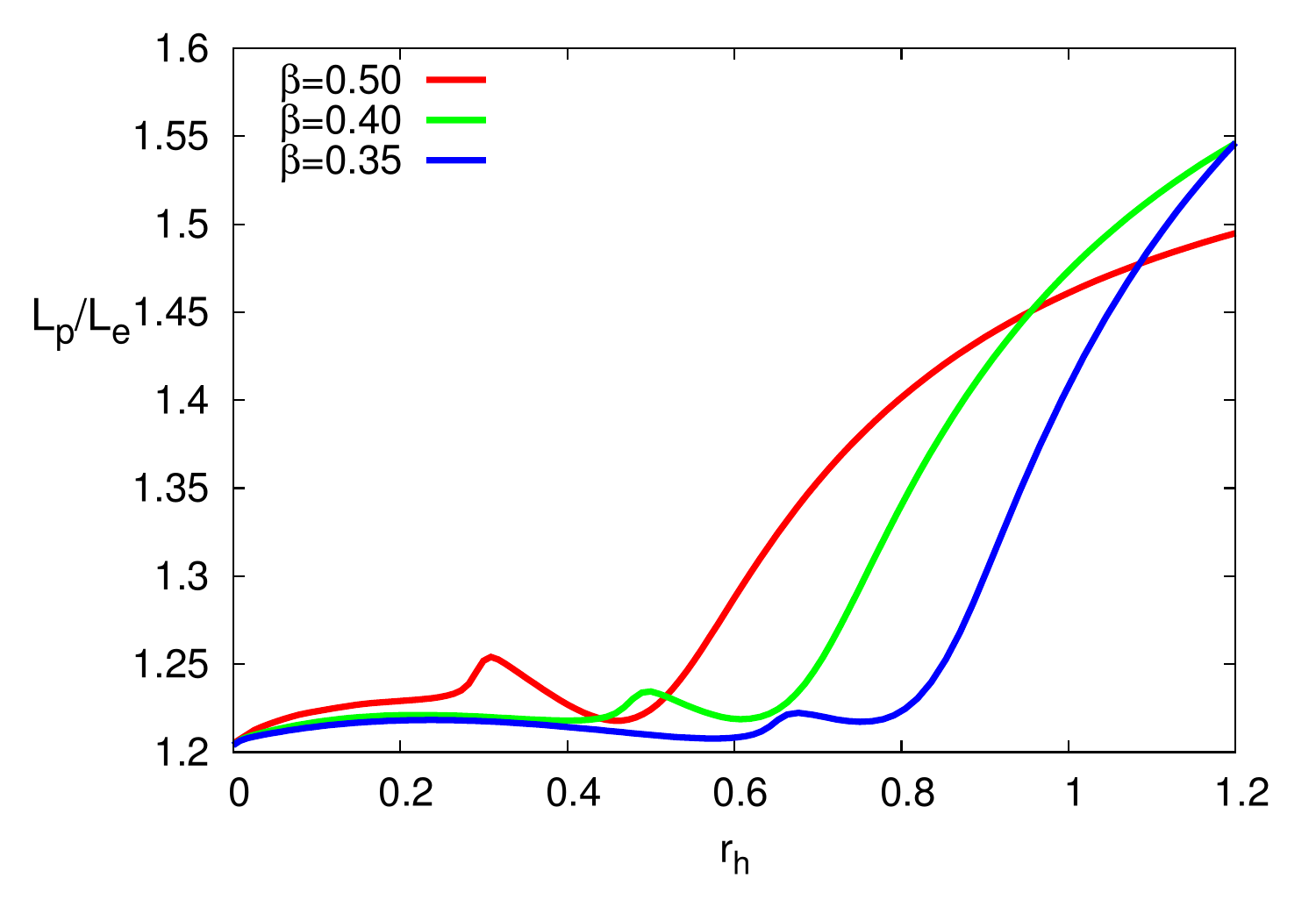}
    \includegraphics[width=0.33\linewidth, angle = -90]{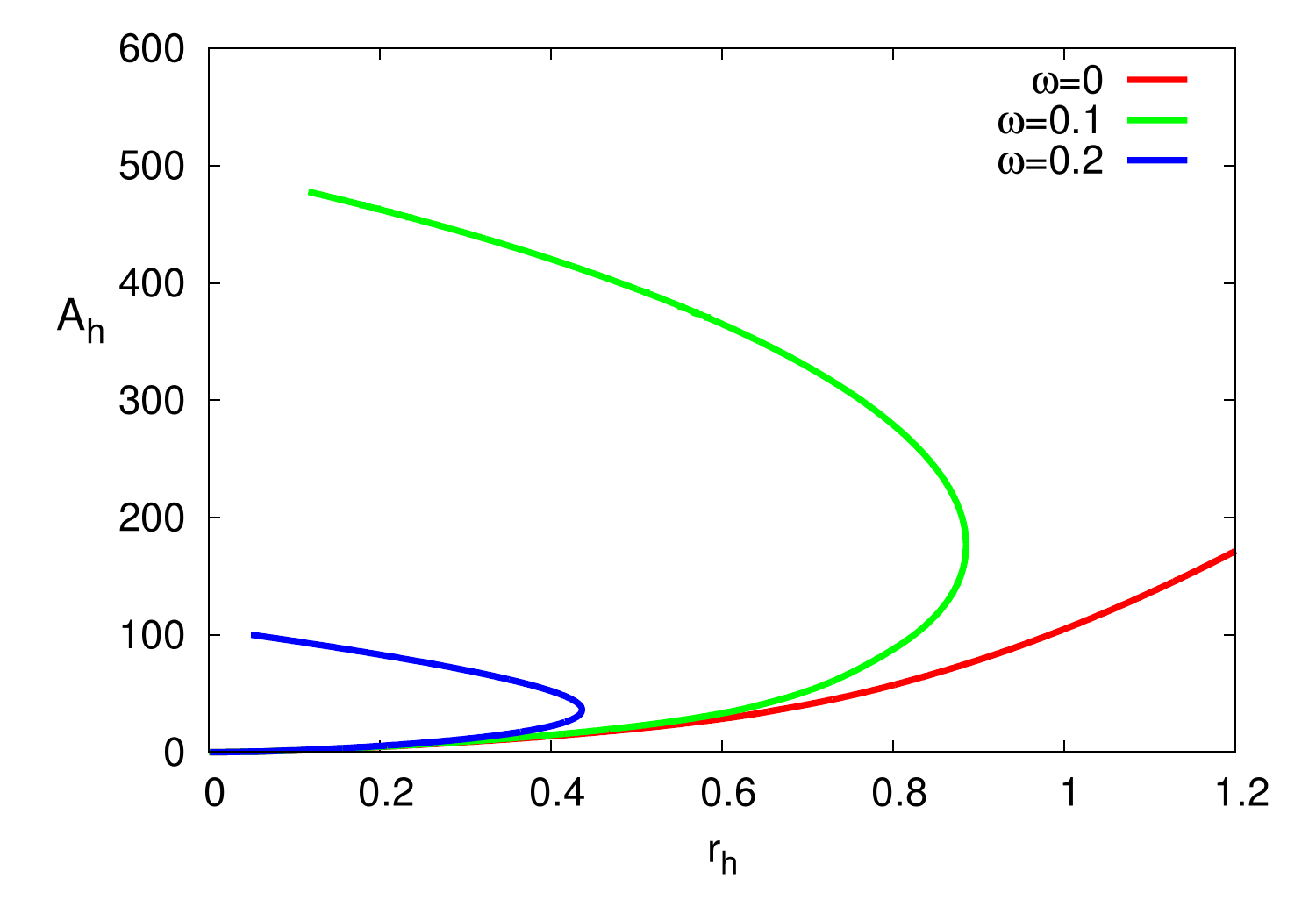}
    \includegraphics[width=0.33\linewidth, angle = -90]{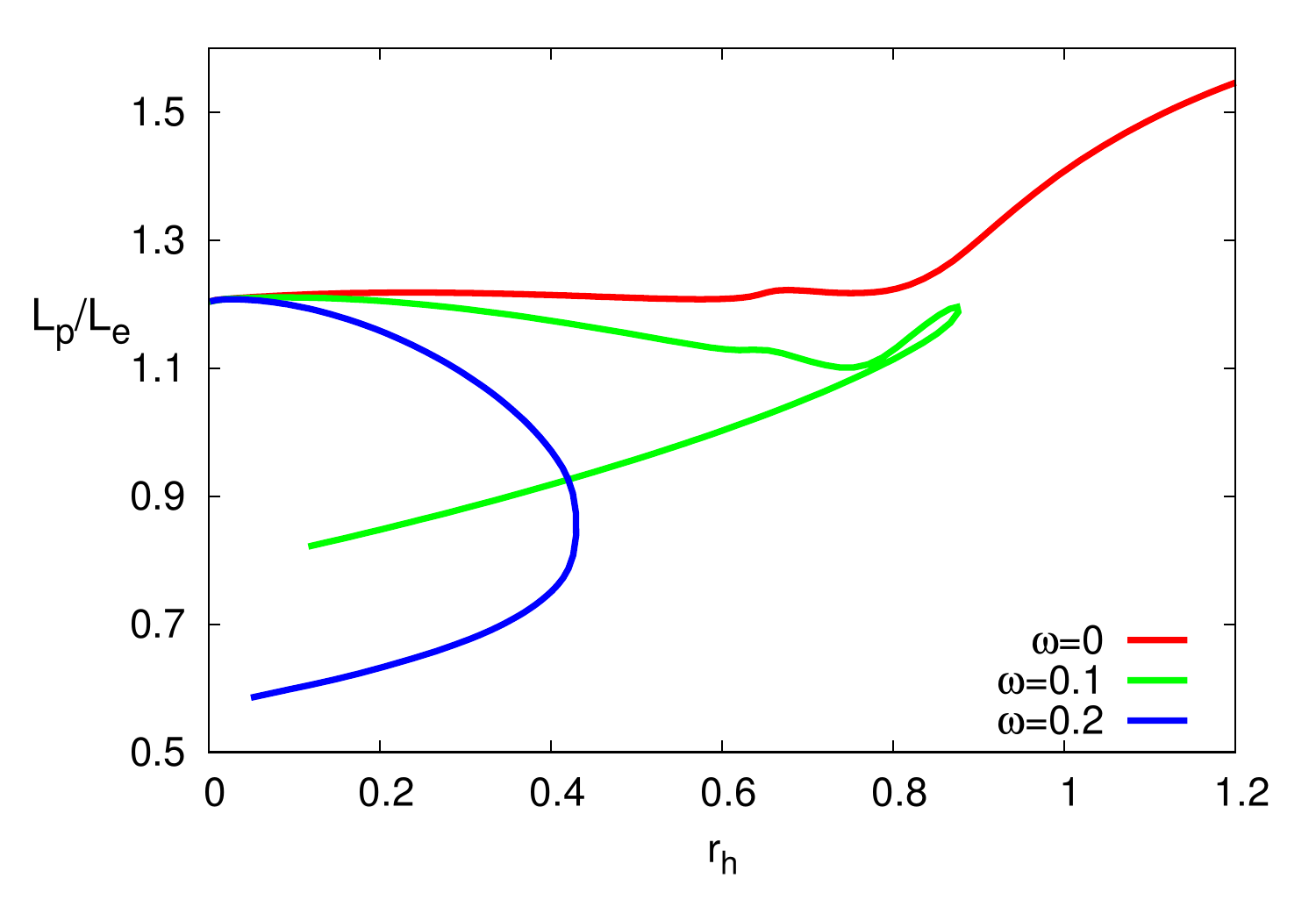}
    \caption{Type II $O(3)$ hairy black holes: the horizon area $A_h$  (left) and ratio of polar to equatorial circumferences (right) versus the horizon radius $r_h$ for $\omega=0$ and several values of $\beta$ (top), and for $\beta=0.35$ and several values of $\omega$ (bottom), $\alpha=0.5$.}
    \label{fig_type23_AEvsRh}
\end{figure}

\begin{figure}[t!]
    \centering
    \includegraphics[width=0.33\linewidth, angle = -90]{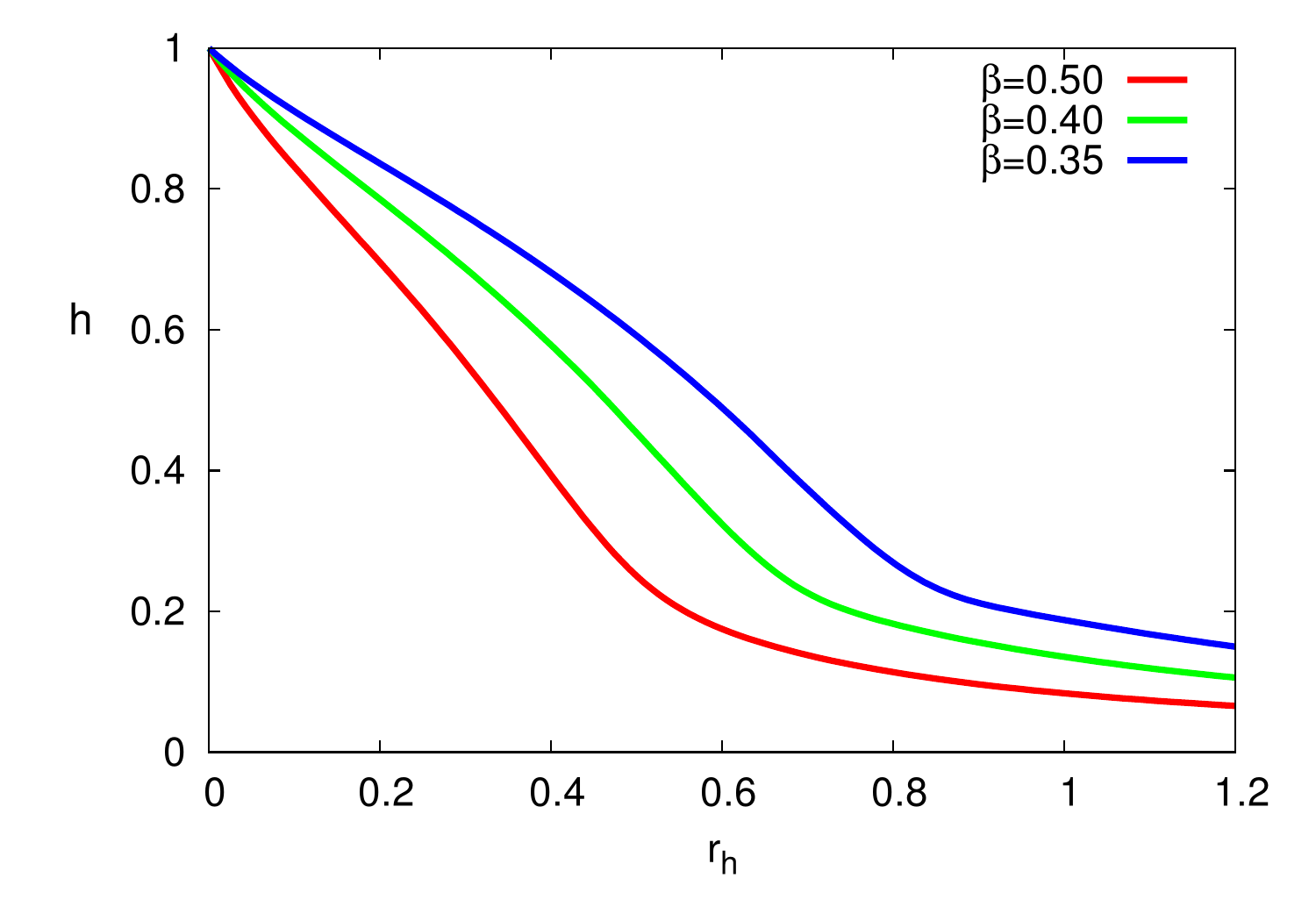}
    \includegraphics[width=0.33\linewidth, angle = -90]{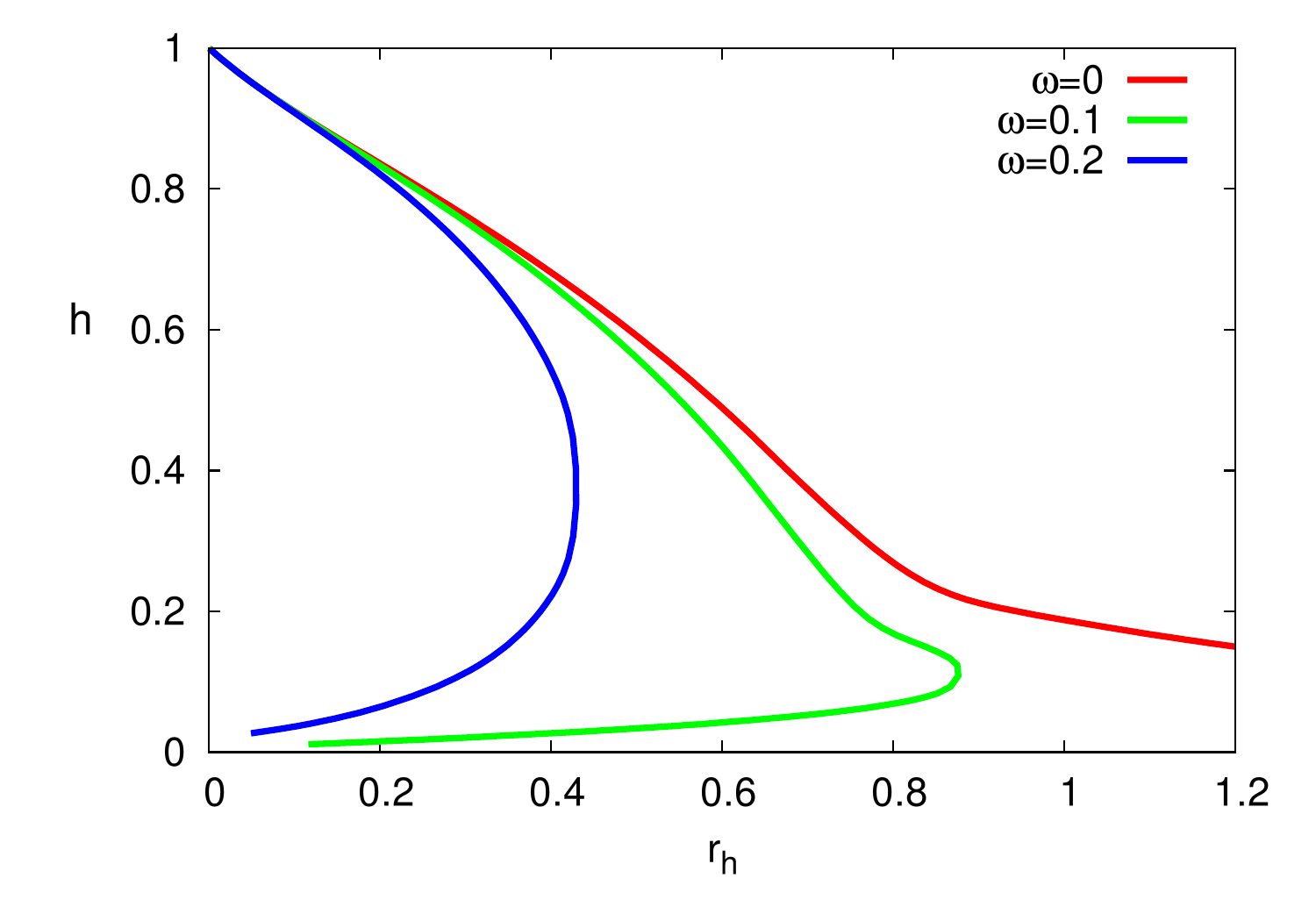}
    \caption{Type II $O(3)$ hairy black holes: the hairiness $h$ versus the horizon radius $r_h$ for $\omega=0$ and some values of $\beta$ (left), and for $\beta=0.35$ and some values of $\omega$ (right); for both $\alpha=0.5$.}
    \label{fig:placeholder}
\end{figure}

Figure~\ref{fig_type23_AEvsRh}, upper plots, illustrate the properties of the  type II non-rotating ($\omega=0$) configurations.
The horizon area $A_h$ (left column) and the eccentricity $L_p/L_e$ (right column) are displayed vs the horizon radius for a few values of $\beta$. 
The horizon area increases monotonically, an increase in the parameter $\beta$ leads to more rapid growth. 
The eccentricity of these Schwarzschild-like hairy black holes is more complex. 
While the standard rotating Kerr-like hairy black holes of type I are strictly oblate ($L_p/L_e < 1$), cf.~Fig.~\ref{fig_type1_AEvsRh}, right plot, the type II non-rotating configurations are always prolate, see Fig.~\ref{fig_type23_AEvsRh}, upper right plot. 
The eccentricity of these configurations depends on the horizon radius in a more complex manner, it initially increases slightly as the horizon radius grows, then decreases, and subsequently rises rapidly with further increases in $r_h$. 
The hairiness of the non-rotating type II configurations decreases monotonically as $r_h$ grows, see Fig.~\ref{fig:placeholder}, left plot.

The emergence of angular momentum as $\omega$ increases from zero changes the situation. 
For a fixed non-zero frequency value, a two-branch pattern emerges in the dependencies of the eccentricity and the horizon area, as displayed in Fig.~\ref{fig_type23_AEvsRh}, bottom plots and Fig.~\ref{fig:placeholder}, right plot. 
Along these branches the horizon area of the type II solutions increases, while the hairiness decreases from one in the solitonic limit ($r_h\to 0$) approaching zero on the backward branch. 
These hairy black holes are initially prolate; however, the eccentricity becomes less than one on the second branch as the hairiness decreases and the configuration approaches the Kerr limit.

%%%%%%%%%%%%%%%%%%%%%%%%%%%%%%%%%%%%%%%%%%%%%%%%%%%%%%%%%%%%%%%%%%%%%
%%%%%%%%%%%%%%%%%%%%%%%%%%%%%%%%%%%%%%%%%%%%%%%%%%%%%%%%%%%%%%%%%%%%%

%%%%%%%%%%%%%%%%%%%%%%%%%%%%%%%%%%%%%%%%%%%%%%%%%%%%%%%%%%%%%%%%%%%%%
%%%%%%%%%%%%%%%%%%%%%%%%%%%%%%%%%%%%%%%%%%%%%%%%%%%%%%%%%%%%%%%%%%%%%

\section{Conclusions and outlook}

Here we have investigated globally regular spinning solitons and stationary axially symmetric black holes with synchronized scalar hair in the gravitating nonlinear O(3) sigma model with a partly negative symmetry-breaking potential. 
For azimuthal winding number $n=1$, we have obtained two qualitatively different families of solutions. 
The type I configurations are continuously connected to the vacuum in the limit where the scalar frequency approaches the mass threshold, whereas the type II configurations belong to the scalaron-supported sector associated with the partly negative potential. 
The corresponding regular spinning solitons exhibit a rich branch structure, including the reconnection between the two families as the gravitational coupling decreases.

These two families persist when an event horizon is introduced. 
The type I black holes resemble previously known black holes with synchronized scalar hair and possess branches that terminate on the Kerr family. 
In contrast, the type II black holes exhibit a considerably different scalar field structure. 
While the scalar field takes its vacuum value on the symmetry axis and at spatial infinity, it may approach the anti-vacuum in an extended region near the equatorial plane. 
Moreover, the type II solutions can possess substantially larger horizon radii than their type I counterparts.

For sufficiently small values of the gravitational coupling $\alpha$, the branch connectivity changes. 
The fundamental branches of type I and type II reconnect to form one continuous family, while the corresponding second branches reconnect to form another. 
The former connects the vacuum-connected type I solutions to the scalaron-supported type II solutions, while the latter emerges from the Kerr family and approaches Kerr-like global properties again as the frequency decreases.

A particularly interesting feature of the type II black holes is their horizon geometry. 
The numerical results reveal significant deviations from the corresponding Kerr geometries, including pronounced prolate horizon deformations. 
Such deformations are also observed in the nonrotating limit. 
While rotation leads to oblate horizon deformations in Kerr black holes, the type II solutions can exhibit pronounced prolate deformations even for nonzero angular velocity. 
The persistence of this effect in the nonrotating limit indicates that it is associated with the anisotropic scalar-field distribution, whose influence can outweigh the rotational flattening of the horizon.
Furthermore, the scalar contribution to the total mass may become relatively small while the deformation of the horizon remains appreciable. 
The mass of the hairy solutions may also approach or cross that of the corresponding Kerr black holes without the scalar field becoming trivial. 
Thus, the global charges alone do not provide a sufficient characterization of the scalar hair or of its influence on the near-horizon geometry.

An important question concerns the dynamical stability of the type I and type II black-hole families, which has not been investigated here. 
The geometry and possible nontrivial topology of the ergosurfaces of the rotating hairy black holes also remain to be investigated.
A further direction is to investigate the existence and properties of self-gravitating topological solitons, Hopfions, in a U(1)-gauged nonlinear O(3) sigma model with a symmetry-breaking potential, minimally coupled to Einstein gravity.
These issues provide interesting directions for future work. 
Overall, our results demonstrate that the partly negative potential of the O(3) sigma model gives rise to a rich family of asymptotically flat black holes with nontrivial scalar hair and horizon geometries that differ substantially from those of the conventional vacuum-connected solutions.

%%%%%%%%%%%%%%%%%%%%%%%%%%%%%%%%%%%%%%%%%%%%%%%%%%%%%%%%%%%%
\section*{Acknowledgements}
We would like to thank Eugeny Babichev, Carlos Herdeiro and Eugen Radu for enlightening discussions. 
Y.S. thanks Carlos Herdeiro and the Gr@v group at the University of Aveiro for their warm hospitality during the completion of this work and
gratefully acknowledges the support from the 
Portuguese Foundation for Science and
Technology (FCT—Fundacão para a Ciência e a
Tecnologia) 
mobility programme, grant RE-C06-i06.M02. 

%%%%%%%%%%%%%%%%%%%%%%%%%%%%%%%%%%%%%%%%%%%%%%%%%%%%%%%%%%%
%\bibliographystyle{abbrv}

%%%%%%%%%%%%%%%%%%%%%%%%%%%%%%%%%%%%%%%%%%%%%%%%%%%%%%%%%%%%%%%%%
\begin{small}

\end{small}
\end{document}